\documentclass[onecolumn,11pt, journal]{IEEEtran}

\usepackage{verbatim,amsmath,enumerate,amssymb}
\usepackage[square,comma,numbers]{natbib} 
\usepackage{inputenc}
\usepackage[T1]{fontenc}
\usepackage{amsmath,amssymb,color,verbatim}

\def\norm#1{\left\| #1 \right\|}
\newtheorem{definition}{Definition}
\newtheorem{thm}{Theorem}
\newtheorem{proposition}{Proposition}
\newtheorem{lemma}{Lemma}
\newtheorem{corollary}{Corollary}
\newtheorem{conjecture}{Conjecture}
\newtheorem{exam}{Example}
\newtheorem{rem}{Remark}
\usepackage{tikz}
\tikzset { domaine/.style 2 args={domain=#1:#2} }
\def\diag{\textrm{diag}}

\DeclareMathOperator*{\tr}{tr}

\providecommand{\abs}[1]{\ensuremath{\left\lvert #1 \right\rvert}}
\providecommand{\norm}[1]{\ensuremath{\left\Vert #1 \right\Vert}}
\providecommand{\bnorm}[1]{\ensuremath{\big\Vert #1 \big \Vert}}

\providecommand{\floor}[1]{\ensuremath{\left\lfloor #1 \right\rfloor}}

\providecommand{\vv}[1]{\textquotedblleft #1\textquotedblright}

\newcommand{\Q}{\mathbb{Q}}
\newcommand{\Z}{\mathbb{Z}}
\newcommand{\C}{\mathbb{C}}
\newcommand{\R}{\mathbb{R}}
\newcommand{\N}{\mathbb{N}}

\newcommand{\ba}{\boldsymbol\alpha}
\renewcommand{\H}{\mathbb{H}}

\DeclareMathOperator*{\out}{out}

\providecommand{\abs}[1]{\ensuremath{\left\lvert #1 \right\rvert}}
\providecommand{\norm}[1]{\ensuremath{\left\Vert #1 \right\Vert}}

\providecommand{\vv}[1]{\textquotedblleft #1\textquotedblright}
\newcommand*{\dotleq}{\mathrel{\dot{\leq}}}
\newcommand*{\dotgeq}{\mathrel{\dot{\geq}}}

\newcommand{\D}{{\mathcal D}}

\newcommand{\A}{{\mathcal A}}

\newcommand{\mindet}[1]{\hbox{\rm det}_{min}\left( #1\right)}

\renewcommand{\IEEEQED}{\IEEEQEDopen}

\begin{document}

\title{The DMT of Real and Quaternionic Lattice Codes and DMT Classification of Division Algebra Codes}

\author{Roope Vehkalahti and Laura Luzzi 

\thanks{%
The research of R. Vehkalahti was supported by the Academy of Finland grant \#299916. \par
This work was presented in part at the IEEE International Symposium on Information Theory (ISIT 2018), Vail, CO \cite{LV18}.
}
\thanks{%
R. Vehkalahti is with the Department of Communications and Networking, FI-02150, Aalto University, Espoo, Finland (e-mail: roope.vehkalahti@aalto.fi). \par
L. Luzzi is with ETIS, UMR 8051  (CY Universit\'{e}, ENSEA, CNRS), 95014 Cergy-Pontoise, France (e-mail: laura.luzzi@ensea.fr).
}
}

\maketitle

\begin{abstract}

In this paper we consider the diversity-multiplexing gain tradeoff (DMT) of so-called minimum delay asymmetric space-time codes. Such codes are less than full dimensional lattices in their natural ambient space. Apart from the multiple input single output (MISO) channel there exist very few methods to analyze the DMT of such codes. Further, apart from the MISO case, no DMT optimal asymmetric codes are known.

We first discuss previous criteria used to analyze the DMT of space-time codes and comment on why these methods fail when applied to asymmetric codes.  We then  consider two special classes of asymmetric codes   where the code-words are restricted to either real or quaternion matrices. We prove two separate diversity-multiplexing gain trade-off (DMT) upper bounds for such codes and provide a criterion for  a lattice code to achieve these upper bounds. We also show that lattice codes based on $\Q$-central division algebras satisfy this optimality criterion.  As a corollary this result provides a DMT classification for all $\Q$-central division algebra codes that are based on standard embeddings.  While the  $\Q$-central division algebra based codes achieve the largest possible DMT of a 
code restricted to either real or quaternion space, they still fall short of the optimal DMT apart from the MISO case.

\end{abstract}


\section{Introduction}

The DMT \cite{ZT}  is a powerful tool for analyzing  the performance of a space-time block code in one shot MIMO communication. Analyzing the DMT curve of a given  code gives us a 
good grasp of the expected performance of the code 
over 
the Rayleigh fading channel. It is therefore of great interest to develop
methods 
to measure the DMT  of a given code.

The previous research reveals that this task is 
non-trivial.
When the diversity-multiplexing gain trade-off was introduced in 2003 in  by Zheng and Tse \cite{ZT}, the only explicit example of a code  achieving the optimal DMT was the Alamouti code \cite{Alamouti}  when it was received with a single antenna. Later in \cite{EKPKL} Elia \emph{et al.} proved
 that the non-vanishing determinant property (NVD) is a sufficient condition 
 for a  $2n^2$-dimensional lattice code in $M_n(\C)$ to achieve the optimal diversity-multiplexing gain trade-off. They also pointed out that division algebra based codes, such as the perfect codes \cite{BORV}, are DMT optimal, and gave a general construction for DMT achieving  $2n^2$-dimensional lattice codes in $M_n(\C)$. This criterion was generalized by 
Tavildar and Viswanath \cite{Tavildar_Viswanath} 
who showed that if the product of  the smallest  $m$ singular values of any non-zero matrix in a $2nm$-dimensional lattice $L\subset M_n(\C)$ stays above some fixed constant, then  $L$ achieves the optimal DMT curve in the $n\times m$ MIMO channel. In the case where $n=m$, this criterion coincides with the NVD condition.

The work in \cite{EKPKL} revealed that there exist $2n^2$-dimensional codes in $M_n(\C)$ achieving the optimal DMT curve, when received with an arbitrary number of receiving antennas $m$. However, in  the scenario where $n>m$, from the decoding complexity point of view, it is in many cases desirable to use  lattice space-time codes that are at maximum  $2nm$-dimensional.
On the other hand, a less than $2nm$-dimensional lattice would be a waste of receiving signal space and energy and automatically leads to a suboptimal DMT curve.  Therefore  a $2nm$-dimensional lattice code is the \vv{best fit} for the $n\times m$ MIMO channel. We refer to such a code as a \emph{well fitting asymmetric space-time code}. In this case
currently the only available criterion for DMT-optimality is the one given in \cite{Tavildar_Viswanath}.

However, 
when $n > m$ asymmetric codes satisfying this condition seem to be very rare. It is also known that there are space-time codes that are DMT optimal despite not satisfying the approximate universality criterion \cite{VeHoLuLa}. 
This motivates the search for a more general and easily applicable 
DMT criterion.

In \cite{SriRa} the authors claimed, when translated into lattice theoretic language,  that any $2nm$-dimensional lattice code $L \subset M_{n}(\C)$ with NVD would achieve the optimal DMT curve 
 with $m$ receive antennas when $n > m$. This 
 would 
imply that  
large families of asymmetric space-time codes are DMT optimal.

In this paper we study the DMT of asymmetric  space-time codes. We begin by reviewing some of the previous DMT criteria and discuss why they seem to fall short when applied to asymmetric codes. We then construct a code that satisfies the DMT optimality  criterion in \cite{SriRa}, but is not DMT optimal. This 
suggests that, unfortunately,  Theorem 2 in  \cite{SriRa} is incorrect. 
Indeed, there are no known DMT optimal asymmetric codes except in the case of MISO channels.

Next, we consider the special class of asymmetric codes 
based on division algebras whose center is $\Q$. This choice seems natural since on one hand, this class includes the Alamouti code \cite{Alamouti}, which is one of the few DMT-optimal asymmetric space-time codes, and on the other hand, 
in  \cite{EKPKL} the optimal codes were based on division algebras. However, the difference is that in \cite{EKPKL} the center of the algebras was complex quadratic, which always leads to lattice codes with full rank $2n^2$ in $M_n(\C)$. 

All the $\Q$-central division algebra codes have the NVD property and several examples
have appeared previously in the
literature \citep{VHO,BelReal,LO, HLL}. However, their DMT was still unknown, apart from Alamouti type codes in the $2\times 1$ channel \cite{ZT}. 

 Unlike 
 the case of complex quadratic center, we 
 show that $\Q$-central division algebras are divided into two categories with respect to their DMT performance. 
This  distinction is  based   on 
the ramification of the infinite Hasse-invariant of the division algebra, which  
determines whether the corresponding lattice code 
can  be embedded into real or quaternionic space.

Our DMT classification holds for any multiplexing gain, extending previous partial results in \cite{VLL2013,ISIT2016_MIMO} which were based on the theory of Lie algebras. We note that the approach used in this paper is quite different and more general. In the spirit of \cite{EKPKL} we are  not 
just considering division algebra codes, but
all space-time codes where the codewords are restricted to the real and quaternionic matrices $M_n(\R)$ or $M_{n/2}(\H)$ respectively. We provide DMT upper bounds for both cases, and prove that  $n^2$-dimensional NVD lattice codes inside $M_n(\R)$ (resp. $M_{n/2}(\H)$) achieve the respective upper bounds. As the $\Q$-central division algebra codes are of this type,
we get their DMT as a corollary.  We note that while these codes achieve the best possible DMT for 
their natural ambient spaces, they don't achieve the general optimal DMT, the only exception being quaternionic codes in the $2\times 1$ channel.

Finally 
we consider the DMT in 
multi-block channels, where we are allowed to encode and decode over a number of independently faded blocks.  Again we find the best possible DMT of asymmetric multi-block codes whose elements belong either to real or quaternionic space and prove that certain division algebra based codes achieve this upper bound. This analysis also provides the DMT  classification of all division algebras whose center is totally real.

\subsection*{Organization of the paper} 
Section II reviews the definition of diversity-multiplexing gain trade-off and basic properties of matrix lattices.  
Section III summarizes previous criteria for DMT-optimality and provides a counterexample to show that the NVD property is not sufficient for DMT-optimality in the asymmetric case. Section IV establishes DMT upper bounds for real and quaternionic space-time codes, and shows that codes with the NVD property achieve these upper bounds. Section V shows how to obtain real and quaternionic lattices with the NVD property from the embeddings of $\Q$-central division algebras, and presents a conjecture about the DMT of space-time codes arising from the regular representations of these algebras. Finally, Section VI extends the results of Section IV to the multi-block case.

\section{Notation and preliminaries} 


\subsection{Single-block channel model and DMT}
Throughout the paper we will consider a MIMO system with $n$ transmit and $m$ receive antennas, and minimal delay $T=n$. 
The received signal is\footnote{A more general multi-block MIMO channel model will be considered in Section \ref{multi-block}.} 
\begin{equation} \label{channel}
Y_c=\sqrt{\frac{\rho}{n}} H_c \bar{X} + W_c,
\end{equation}
where $\bar{X} \in M_n(\C)$ is the transmitted codeword, $H_c \in M_{m,n}(\C)$ and $W_c \in M_{m,n}(\C)$ are the channel and noise matrices with i.i.d. circularly symmetric complex Gaussian entries $h_{ij}, w_{ij} \sim \mathcal{N}_{\C}(0,1)$, and $\rho$ is the signal-to-noise ratio (SNR). 
We suppose that perfect channel state information is available at the receiver but not at the transmitter, and that maximum likelihood decoding is performed. 

\begin{definition}\label{def:stbc}
A {\em space-time block code} (STBC) $C$  for some designated  SNR level $\rho$ is a set of $n \times n$ complex matrices satisfying the average power constraint
\begin{equation}\label{power_constraint}
\frac{1}{\abs{C}}\sum_{X \in C} \norm{X}_F^2 \leq  n^2 . 
\end{equation}
 A coding scheme $\{ C(\rho)\}$ is a family of STBCs, one for each SNR level. The rate for the code $C(\rho)$ is $R(\rho)=\frac{1}{T} \log \abs{C(\rho)}$. 
\end{definition}

 We say that the coding scheme $\{C(\rho)\}$ achieves the \emph{diversity-multiplexing gain trade-off} (DMT) of \emph{spatial multiplexing gain} $r$ and \emph{diversity gain} $d(r)$ if the rate satisfies
\begin{equation}\label{ratedemand}
\lim_{\rho \to \infty} \frac{R(\rho)}{\log(\rho)} = r,
\end{equation}
and the average error probability is such that
\[
P_e(\rho) \ \doteq \ \rho^{-d(r)},
\]
where by the dotted equality we mean $f(M) \doteq g(M)$ if 
\begin{equation}
\lim_{M\to \infty}\frac{\log(f(M))}{\log(M)} = \lim_{M\to \infty}\frac{\log(g(M))}{\log(M)}. \label{eq:dotdefn}
\end{equation}
Notations such as $\dot\geq$ and $\dot\leq$ are defined in a similar way.

With the above definitions, the main result in \cite{ZT} is the following.
\begin{thm}[Optimal DMT] \label{thm:DMT}
Let $n$, $m$, $T$, $\{C(\rho)\}$, and $d(r)$ be defined as before. Then any STBC coding scheme $\{C(\rho)\}$ has error probability lower bounded by
\begin{equation}
P_e(\rho) \ \dot\geq\ \rho^{-d^*(r)} \label{eq:DMT1}
\end{equation}
or equivalently, the diversity gain
\begin{equation}
d(r) \leq d^*(r), \label{eq:DMT2}
\end{equation}
when the coding is limited within a block of $T$ channel uses. The optimal diversity gain $r \mapsto d^*(r)$, also termed the optimal DMT, is a piece-wise linear function connecting the points $(r,(n-r)(m-r))$ for $r=0,1,\ldots,\min\{n,m\}$. 
\end{thm}

\subsection{Matrix Lattices and their coding schemes}\label{basic}
  
  In this section we describe how to obtain a coding scheme  that satisfies the rate condition \eqref{ratedemand} and average energy condition \eqref{power_constraint} from  a matrix lattice $\mathcal{L} \subseteq M_n(\C)$.

\begin{definition}\label{def:lattice}
A {\em matrix lattice} $\mathcal{L} \subset M_n(\C)$ has the form
$$
\mathcal{L}=\Z B_1\oplus \Z B_2\oplus \cdots \oplus \Z B_k,
$$
where the matrices $B_1,\dots, B_k$ are linearly independent over $\R$, i.e., form a lattice basis, and $k$ is
called the \emph{rank}  or the \emph{dimension} of the lattice.
\end{definition}

\begin{definition}\label{def:NVD}
If the minimum determinant of the lattice $\mathcal{L} \subset M_n(\C)$ is non-zero, i.e. it satisfies
\[
\inf_{{\bf 0} \neq X \in \mathcal{L}} \abs{\det (X)} > 0, 
\]
we say that the lattice satisfies the \emph{non-vanishing determinant} (NVD) property.
\end{definition}

Let $\norm{X}_F = \sqrt{\tr(X^\dagger X)}$ denote the Frobenius norm of $X$.

\begin{definition}[Spherical shaping]
Given a positive real number $M$ and a $k$-dimensional lattice $\mathcal{L} \subset M_{n}(\C)$, we define
$$
\mathcal{L}(M)=\{X \in \mathcal{L} \;:\; \norm{X}_F \leq M,\; X\neq {\bf 0} \}.
$$
\end{definition}



The following two results are well known \cite{Kratzel}.
\begin{lemma}\label{spherical}
If $\mathcal{L}$ is a  $k$-dimensional lattice in  $M_{n}(\C)$ and
$\mathcal{L}(M)$ is defined as above, 
then
\[
|\mathcal{L}(M)|= cM^{k}+ O(M^{k-1}),
\]
where $c$ is some  positive  constant, independent of $M$. 
\end{lemma}

In particular, it follows that we can choose real constants $K_1$ and $K_2$ such that
\begin{equation}\label{latticepointbound}
K_1M^k\geq |\mathcal{L}(M)|\geq K_2 M^k.
\end{equation}

\begin{lemma}\label{Mclaurin}
Let $\mathcal{L}$ be a $k$-dimensional lattice in $M_n(\C)$. Then
\begin{eqnarray*}
 s_2 M^{k+2}\leq \sum_{X\in \mathcal{L}(M)} \norm{X}_F^2\leq s_1M^{k+2},
\end{eqnarray*}
where $s_1$ and $s_2$ are constants independent of $M$.
\end{lemma}

We can now give a formal definition of a family of space-time lattice codes of finite size.
\begin{definition}
Given the lattice $\mathcal{L} \subset M_n(\C)$, a space-time lattice coding scheme associated with $\mathcal{L}$ is a collection of STBCs given by
\begin{equation}\label{codingscheme}
C_{\mathcal{L}}(\rho)=\rho^{ -\frac{rn}{k}}\mathcal{L}\left(\rho^{\frac{rn}{k}}\right)
\end{equation}
for the desired multiplexing gain $r$ and for each $\rho$ level.
\end{definition}
One can see that according to Lemma \ref{spherical} the coding scheme defined this way indeed has multiplexing gain $r$.

From  Lemma \ref{Mclaurin} we have
$$
 \sum_{X \in \mathcal{L}\left(\rho^{\frac{rn}{k}}\right)} \rho^{-\frac{2rn}{k}} \norm{X}_F^2 \doteq \rho^{-\frac{2rn}{k}}(\rho^{\frac{rn}{k}})^{k+2}=\rho^{rn}.
$$
On the other hand we also have that $|\mathcal{L}(\rho^{\frac{rn}{k}})| \doteq\rho^{rn}$ from Lemma \ref{spherical}. Combining the above shows that the code  $C_{\mathcal{L}}( \rho)$ has the correct average power \eqref{power_constraint} from the DMT perspective, i.e., in terms of the dotted equality. 

\begin{rem}
We discussed the question of transforming a lattice code into a coding scheme in detail since in Section \ref{failing} we will prove that a certain lattice code is not DMT optimal. It is therefore crucial that our coding schemes are using the lattices in an asymptotically optimal way. 

\end{rem}


\section{Previous criteria for DMT optimality and failing of the NVD condition}
Several methods have been proposed to analyze the DMT of a space-time code, but most of them are not tight enough to prove DMT-optimality except for special cases. 
For example, in \cite{ZT} the authors analysed the  DMT of different versions of BLAST \cite{blast}. They also showed the DMT optimality of the Alamouti code over the $2\times 1$ channel by transforming the MISO channel into two parallel channels. A similar approach was used to prove that different diagonal space-time codes are DMT-optimal \cite{Tavildar_Viswanath}. However, this criterion can be only applied to special classes of codes.

Using the union bound for the error probability to evaluate the DMT \cite{Tavildar_Viswanath}  is a universal approach that can be used to analyze any kind of space-time codes. However, it consistently gets too loose when the multiplexing gain is high \cite{YW, ISIT2016_MIMO}.

So far the most effective criterion to prove DMT optimality is the NVD criterion \cite{YW, EKPKL}. This criterion was generalized by Tavildar and Viswanath in \cite{Tavildar_Viswanath}.  We begin by shortly reviewing their approximate universality (AU) criterion and draw some implications of their work for the lattice based coding schemes introduced in the previous section.  We note that  AU is a  considerably stronger condition that 
implies DMT. In particular a space-time code can be DMT optimal despite not being approximately universal.   

\begin{thm}\label{Visva}
A sequence of codes $C(\rho)$ of rate $R(\rho)$  is approximately universal over the $n\times m$ MIMO channel if and only if, for every pair of distinct codewords $X, \bar{X} \in C(\rho)$,
\begin{equation}
\lambda_{1}^2\cdots\lambda_s^2\geq\frac{1}{2^{R(\rho) + o(\log{\rho})}},
\end{equation}
where $\lambda_1,\ldots,\lambda_s$ are the smallest $s$ singular values of the codeword difference matrix $X-\bar{X}$ and 
$s=\min(m, n)$.
\end{thm}

Here the notation $o(\log{\rho})$ refers to a function that is  asymptotically  dominated by $\epsilon \log{\rho}$ for any $\epsilon>0$. \\
In the case $m\geq n$, this condition is simply the NVD condition of Definition \ref{def:NVD}.

\begin{definition}
We refer to the $i$-th smallest singular value of the matrix $X$ with $\lambda_i(X)$
and for $s \leq n$, we set 
$$
 \Delta_s(X)=\prod_{i=1}^{s}\lambda_i^2(X).
$$
\end{definition}
We can now extend this definition to lattices.
\begin{definition}
Given a lattice $\mathcal{L} \subset M_n(\C)$, we define
$$
\Delta_s(\mathcal{L}):=\mathrm{inf}\{\Delta_s(X)\mid X\in \mathcal{L}  \setminus \{0\} \}.
$$
\end{definition}

The result by Tavildar and Viswanath now transforms into the following.
\begin{corollary}\label{NVD2}
Suppose that $n \geq m$, $\mathcal{L}$ is a $2mn$-dimensional lattice 
in $M_n(\C)$ and that 
$$
\Delta_{m}(\mathcal{L})\neq 0.
$$
Then $C_{\mathcal{L}}(\rho)$ is approximately universal (and therefore DMT optimal), when received with $m$ antennas.
\end{corollary}
\begin{IEEEproof}
Assume without loss of generality that we have scaled our lattice so that $\Delta_{m}(\mathcal{L})=1$.
The finite codes we consider are of the type $C_{\mathcal{L}}(\rho)=\rho^{-\frac{r}{2m}}\mathcal{L}(\rho^{\frac{r}{2m}})$.
Given two codewords $\rho^{-\frac{r}{2m}}X$ and $\rho^{-\frac{r}{2m}}\bar{X}$ in $\rho^{-\frac{r}{2m}}\mathcal{L}(\rho^{\frac{r}{2m}})$,  we  have 
$$
\Delta_{m}\left(\rho^{-\frac{r}{2m}}\left(X-\bar{X} \right)\right)=\rho^{-r}\Delta_{m}(X-\bar{X})\geq \rho^{-r}.
$$
The last inequality here follows as $X-\bar{X}\in \mathcal{L}$ and we assumed that $\Delta_{m}(\mathcal{L})=1$.
On the other hand according to equation \eqref{latticepointbound}  we have $\abs{C_{\mathcal{L}}(\rho)} \doteq \rho^{rn}$ and 
$$\frac{1}{2^{R(C_{\mathcal{L}}(\rho))}} \leq \frac{1}{2^{\log(A\rho^r)}}=\frac{1}{A\rho^r},$$ for some constant $A$ independent of $\rho$. We obviously  have that $A\in 2^{o(\log(\rho))}$. \end{IEEEproof}

\begin{rem}
The reader should note that approximate universality does allow a vanishing product of singular values for a lattice code.
However, this vanishing must be in the class $2^{-o(\log(\rho))}$. In particular vanishing with speed $\rho^{-\epsilon}$ is not allowed, for any fixed 
$\epsilon$.   For simplicity we stated Corollary \ref{NVD2}  in a more demanding form where we require the non-vanishing condition and not just the dotted version.
\end{rem}

\begin{exam}
The Alamouti code together with QAM modulation can be seen as a $4$-dimensional lattice  code $\mathcal{L}_{\text{Alam}}\subset M_2(\C)$. For this code 
$\Delta_1(\mathcal{L})>0$. 
Therefore the coding scheme $C_{\mathcal{L}_{\text{Alam}}}(\rho)$ is approximately universal when received with a single antenna.
\end{exam}

\begin{exam}
The division algebra based codes such as the  Perfect codes \cite{BORV}  are $2n^2$-dimensional lattices in 
$M_n(\C)$ and have the NVD property and are therefore DMT optimal.
\end{exam}

However, the conditions of Corollary \ref{NVD2} seem difficult to satisfy in other cases. 
As a matter of fact we conjecture the following:
\begin{conjecture}
The conditions  of Corollary \ref{NVD2}  can be satisfied only when either $m=n$ or when $n=2$ and  $m=1$.
\end{conjecture}

\subsection{Failing of the NVD criterion}\label{failing}
Many codes are DMT optimal despite not satisfying the approximate universality criterion of the previous section.
 For example the diagonal number field codes \cite{DamenBelfi} and many of the fully diverse quasi-orthogonal codes \cite{SuXia}  are DMT optimal in the $n\times 1$ MIMO channel \cite{VeHoLuLa}. Seen as lattice codes, these are $2n$-dimensional lattices  in $M_n(\C)$ and have the NVD property.  However, they are not approximately universal \cite{EJ}.

It is a tempting idea that the NVD condition for   a $2nm$-dimensional lattice $\mathcal{L} \subseteq M_n(\C)$ would be enough for the coding scheme $C_{\mathcal{L}}(\rho)$ to be DMT optimal when received with $m$ receiving antennas. This  was suggested in \cite{SriRa}.

Using the normalization in \cite{SriRa}  we can state the NVD   condition for a $2nm$ dimensional lattice $\mathcal{L}$ and scheme $\rho^{\frac{1}{2}}C_{\mathcal{L}}(\rho)=\rho^{\frac{1}{2} -\frac{r}{2m}}\mathcal{L}\left(\rho^{\frac{r}{2m}}\right)$  in the form
\begin{equation}\label{weakNVD}
\Delta_n(X)\geq c\rho^{n(1- \frac{r}{m})},
\end{equation}
for any non-zero codeword $X$ in $\rho^{1/2}C_{\mathcal{L}}(\rho)$ and fixed  positive constant $c$. According to  Theorem 2 in  \cite{SriRa} this should be a sufficient condition for achieving the optimal DMT.

However, this is not the case and we will   now build a code for the  $4\times 1$ MISO channel  that  satisfies the criterion
\eqref{weakNVD}, but is not DMT optimal in this channel.

\begin{rem}
One should notice that condition \eqref{weakNVD} is considerably weaker than the condition in Corollary \ref{NVD2}. 
Using the normalization of  \cite{SriRa}, the condition of Corollary \ref{NVD2} can be written as follows:
if  a coding scheme $\rho^{1/2}C_{\mathcal{L}}(\rho)$,  based on a $2mn$-dimensional lattice code $\mathcal{L}$, satisfies
$$
\Delta_{m}(X)\geq c\rho^{m(1- \frac{r}{m})},
$$
for any non-zero codeword $X\in\rho^{1/2} C_{\mathcal{L}}(\rho)$, any $\rho$ and some fixed constant $c$, then 
it 
is approximately universal.
\end{rem}

Let us begin with the Golden Code $\mathcal{L}_{\text{Gold}}$ \cite{RBV}. One can see it as an $8$-dimensional NVD lattice in $M_2(\C)$. 
 According to \eqref{codingscheme} we can  use scheme $\rho^{\frac{1}{2}-\frac{r}{4}}\mathcal{L}_{\text{Gold}}(\rho^{\frac{r}{4}})$ to study the DMT of $\mathcal{L}_{\text{Gold}}$. It was already proven in \cite{EKPKL} that this scheme achieves the optimal DMT curve in the $2\times 2$ MIMO channel.

Let's now transform the Golden Code into an $8$-dimensional code in $M_4(\C)$ by setting
$$
\mathrm{diag}(X,X)=
\begin{pmatrix}
X& \bf{0}\\
\bf{0} &X
\end{pmatrix},
$$
where $X \in M_2(\C)$ and $\bf{0}$ is the $2\times2$ zero matrix. The set $\mathrm{diag}(\mathcal{L}_{\text{Gold}})=\{\mathrm{diag}(X)\,|\, X\in \mathcal{L}_{\text{Gold}}\}$ is an 8-dimensional  NVD lattice code in $M_4(\C)$. In order to satisfy the energy normalization demands we have to consider the
scheme  $\rho^{\frac{1}{2}-\frac{r}{2}}\mathrm{diag}(\mathcal{L}_{\text{Gold}})(\rho^{\frac{r}{2}})=C_{\mathrm{diag}(\mathcal{L}_{\text{Gold}})}(\rho)$.
\begin{proposition}\label{example}
The scheme
$C_{\mathrm{diag}(\mathcal{L}_{\text{Gold}})}(\rho)$ is not a DMT optimal code over the $4\times 1$ MISO channel.
\end{proposition}
\begin{IEEEproof}
Suppose that we transmit a codeword $\mathrm{diag}(X)$, where 
$$
X=
\begin{pmatrix}
x_1& x_2\\
x_3& x_4
\end{pmatrix}.
$$
Given the channel vector $\mathbf{h}=[h_1, h_2, h_3, h_4]$ and the noise $\mathbf{w}=[w_1,w_2,w_3,w_4]$, the received signal is
$$
\mathbf{y}=[y_1, y_2, y_3, y_4] =\mathbf{h}\cdot \mathrm{diag}(X) +\mathbf{w}=
[h_1x_1+h_2x_3, h_1x_2+h_2x_4, h_3x_1+ h_4x_3, h_3x_2+ h_4x_4] +\mathbf{w}.
$$
But this system is equivalent to
$$
\begin{pmatrix}
y_1& y_2\\
y_3& y_4
\end{pmatrix}=
\begin{pmatrix}
h_1& h_2\\
h_3& h_4
\end{pmatrix}
\begin{pmatrix}
x_1& x_2\\
x_3& x_4
\end{pmatrix}
+
\begin{pmatrix}
w_1& w_2\\
w_3& w_4
\end{pmatrix}.
$$
We can see that the error performance of  $ \mathrm{diag}(\mathcal{L}_{\text{Gold}})$ when received with a single antenna is 
exactly that of $\mathcal{L}_{\text{Gold}}$ when received with two antennas.  The DMT for  the coding scheme $\rho^{\frac{1}{2}-\frac{r}{4}}\mathcal{L}_{\text{Gold}}(\rho^{\frac{r}{4}})$ is  
is the piecewise linear function connecting the points $[r,(2-r)(2-r)^+]$ for integer values. However, this is not directly the DMT for $\rho^{\frac{1}{2}-\frac{r}{2}}\mathrm{diag}(\mathcal{L}_{\text{Gold}})(\rho^{\frac{r}{2}})$. This is due to the fact that for 
the diagonal scheme
we have $T=4$ and therefore the diversity gain achieved with multiplexing gain $r$ in the $4\times 1$ channel corresponds to  diversity gain $d(2r)$ in the $2\times 2$ channel. We then see that the DMT of $\rho^{\frac{1}{2}-\frac{r}{2}}\mathrm{diag}(\mathcal{L}_{\text{Gold}})(\rho^{\frac{r}{2}})$
is represented by a line connecting points   $[r,(2-2r)(2-2r)^+]$, where $r=0,\frac{1}{2}, 1$. On the other hand the DMT of the $4\times 1$ MISO channel is simply a straight line between $[0,4]$ and $[1,0]$. 
\end{IEEEproof}

This result shows that 
for a  lattice $\mathcal{L}\subset M_n(\C)$ of dimension smaller than $2n^2$ the NVD condition is not enough for the code to reach the optimal DMT.  

\begin{rem}
We point out that while our counterexample involves coding schemes of the form \eqref{codingscheme}, it generalizes to other schemes.
\end{rem}


\section{The DMT of Real and Quaternion Space-Time Codes}
 In the previous sections we have seen that characterizing the DMT of asymmetric codes is a difficult task.
 In the rest of the paper we propose a new approach that applies to a large class of asymmetric codes. We will 
prove that if the codewords of the space-time scheme belong to a certain restricted set of matrices, its DMT is automatically upper bounded by a limit that is 
tighter than the general DMT bound. We then show that if  the space-time code belongs to this class of codes, has suitable degree and satisfies the NVD condition, it achieves this restricted DMT.  Later, in Section \ref{division_algebra}, we show that codes satisfying these conditions can be obtained from division algebras, and conclude that our DMT upper bounds are tight.

The asymmetric space-time codes  we are considering  live in the subspaces  of  the $2n^2$-dimensional real vector space $M_n(\C)$.  The first such subspace consists of all the  real matrices  inside $M_n(\C)$ and we denote it with $M_n(\R)$. The other subspace of interest consists of quaternionic matrices.

Let us assume that $2\mid n$. We denote  with  $M_{n/2}(\H)$ the set of quaternionic matrices 
$$
\begin{pmatrix}
A &   -B^* \\
B&  A^*                                               
\end{pmatrix}
\in M_{n}(\C),
$$
where $*$ refers to complex conjugation and $A$ and $B$ are complex matrices in $M_{n/2}(\C)$. 

The spaces   $M_{n/2}(\H)$ and $M_n(\R)$ are $n^2$-dimensional real subspaces of $M_n(\C)$.
It follows that if a lattice $\mathcal{L}$ is a subset of either of these subspaces, its dimension is at most $n^2$.

\subsection{Equivalent channel model for real lattice codes}
In this section, we focus on the special case where $\mathcal{C}(\rho) \subset M_n(\R)$, i.e. the code is a set of real matrices. 

First, we show that the channel model (\ref{channel}) is equivalent to a real channel with $n$ transmit and $2m$ receive antennas.\\
We can write $H_c=H_r + i H_i$, $W_c=W_r + i W_i$, where $H_r,H_i,W_r,W_i$ have i.i.d. real Gaussian entries with variance $1/2$. If $Y_c=Y_r + i Y_i$, with $Y_r, Y_i \in M_{m \times n}(\R)$, we can write an equivalent real system with $2m$ receive antennas:
\begin{equation} \label{real_channel}
Y=\begin{pmatrix} Y_r \\ Y_i \end{pmatrix} = 
\sqrt{\frac{\rho}{n}} \begin{pmatrix} H_r \\ H_i \end{pmatrix} \bar{X} + \begin{pmatrix} W_r \\ W_i \end{pmatrix} =\sqrt{\frac{\rho}{n}} H\bar{X} + W,
\end{equation}
where $H \in M_{2m \times n}(\R)$, $W \in M_{2m \times n}(\R)$ have real i.i.d. Gaussian entries with variance $1/2$.

\subsection{General DMT upper bound for real codes} \label{real_upper_bound}
Using the equivalent real channel, we can now establish a general upper bound for the DMT of real codes.

\begin{thm} \label{theorem_real_upper}
Suppose that $\forall \rho$, $\mathcal{C}(\rho) \subset M_n(\R)$. Then the DMT of the code $\mathcal{C}$ is upper bounded by the function $d_1(r)$ connecting the points $(r,[(m-r)(n-2r)]^+)$ where $2r \in \Z$.
\end{thm}

\begin{IEEEproof} This part of the proof closely follows \cite{ZT}. Given a rate $R=r \log \rho$, consider the outage probability \cite{Telatar}
\begin{equation} \label{P_out}
P_{\out}(R)=\inf_{Q \succ 0, \;\tr(Q) \leq n} \mathbb{P}\left\{ \Psi(Q,H) \leq R\right\},
\end{equation}
where $\Psi(Q,H)$ is the maximum mutual information per channel use of the real MIMO channel (\ref{real_channel}) with fixed $H$ and real input with fixed covariance matrix $Q$.\footnote{Unlike \cite{Telatar} and \cite{ZT}, we don't use a strict inequality in the definition (\ref{P_out}), but our definition is equivalent since the set of $H$ such that $\Psi(Q,H)=R$ has measure zero.} 
  Following a similar reasoning as in \cite[Section 3.2]{Telatar}, we have
$$\Psi(Q,H)=\frac{1}{2} \log \det \left(I + \frac{\rho}{n} H Q H^T\right).$$ 
As in \cite[Section III.B]{ZT}, since $\log \det$ is increasing on the cone of positive definite symmetric matrices, for all $Q$ such that $\tr(Q)\leq n$ we have $\frac{Q}{n}\preceq I$ and
$$P_{\out}(R) \geq \mathbb{P} \left\{\frac{1}{2} \log \det (I + \rho HH^T)  \leq R\right\}.$$
Note that $\det(I+\rho HH^T)=\det(I+\rho H^T H)$. Let $L=\min(2m,n)$, and $\Delta=\abs{n-2m}$. Let $\lambda_1 \geq \lambda_2 \geq \cdots \geq \lambda_L > 0$ be the nonzero eigenvalues of $H^T H$. The joint probability distribution of  $\boldsymbol\lambda=(\lambda_1,\ldots,\lambda_L)$ is given by \cite{Edelman}\footnote{We have slightly modified the expression to be consistent with our notation. In \cite{Edelman}, the author considers a matrix $AA^T$ where each element of $A$ is $\mathcal{N}(0,1)$.}:
\begin{equation} \label{p_lambda_real}
p(\boldsymbol\lambda)=Ke^{-\sum\limits_{i=1}^L \lambda_i} \prod_{i=1}^L \lambda_i^{\frac{\Delta-1}{2}}\prod_{i<j}(\lambda_i-\lambda_j)
\end{equation}
for some constant $K$.
 Consider the change of variables $\lambda_i=\rho^{-\alpha_i} \;\forall i$.  The corresponding distribution for $\boldsymbol\alpha=(\alpha_1,\ldots,\alpha_L)$ in the set  $\mathcal{A}=\{\ba\;:\; \alpha_1 \leq \cdots \leq \alpha_L\}$ is
\begin{equation} \label{p_alpha_real}
\!p(\boldsymbol\alpha)\!=\!K(\log \rho)^L e^{-\!\!\sum\limits_{i=1}^L \! \rho^{-\alpha_i}}\!\rho^{-\!\!\sum\limits_{i=1}^L\!\alpha_i \left(\frac{\Delta+1}{2}\right)}\!\prod_{i<j}\!\left(\rho^{-\alpha_i}\!\!-\!\rho^{-\alpha_j}\!\right) 
\end{equation}
Then we have
{\allowdisplaybreaks
\begin{align*}
&P_{\out}(R) \doteq \mathbb{P}\left\{ \prod_{i=1}^L (1+ \rho \lambda_i)  \leq \rho^{2r}\right\}
=\mathbb{P}\left\{ \prod_{i=1}^L(1+\rho^{1-\alpha_i}) \leq \rho^{2r}\right\}.
\end{align*}
}%
To simplify notation, we take $s=2r$. Note that $1+\rho^{1-\alpha_i} \leq 2 \rho^{(1-\alpha_i)^+} \doteq \rho^{(1-\alpha_i)^+}$, therefore
{\allowdisplaybreaks
\begin{align*}
& P_{\out}(R) \dotgeq \mathbb{P}\left\{ \prod_{i=1}^L \rho^{(1-\alpha_i)^{+}} \leq  \rho^s\right\}
\geq \mathbb{P}(\mathcal{A}_0),
\end{align*}
}%
where
{
\begin{align} \label{A_0} 
&\mathcal{A}_0=
\left\{\ba \in \mathcal{A}:\;\alpha_i \geq 0 \;\forall i=1,\ldots,L,\; \sum_{i=1}^L (1-\alpha_i)^+  \leq s\right\} \notag\\
&=\!\left\{\!\ba \in \mathcal{A}\!:\;\!\alpha_j\geq 0,\;\sum_{i=1}^j (1-\alpha_i) \leq s \; \forall j=1,\ldots,L\right\}.
\end{align}
}%
In fact, given $\ba \in \mathcal{A}$, let $t=t(\ba)$ be such that $\alpha_{t+1} \geq 1 \geq \alpha_t$. Then $\forall j=1,\ldots,L$,
$\sum_{i=1}^j (1-\alpha_i) \leq \sum_{i=1}^t (1-\alpha_i)=\sum_{i=1}^L (1-\alpha_i)^+.$\\
Consider $S_{\delta}=\{\ba \in \mathcal{A}:\; \abs{\alpha_i-\alpha_j}> \delta \; \forall i \neq j\}$. Then 
{\allowdisplaybreaks
\begin{align*}
&P_{\out}(R) \dotgeq \int_{\mathcal{A}_0} e^{-\sum\limits_{i=1}^L \rho^{-\alpha_i}} \rho^{- \sum\limits_{i=1}^L\frac{(\Delta+1)\alpha_i}{2}} \prod_{i<j} (\rho^{-\alpha_i}-\rho^{-\alpha_j}) d \ba\\
&\geq \int_{\mathcal{A}_0 \cap S_{\delta}}e^{-\sum\limits_{i=1}^L \rho^{-\alpha_i}} \rho^{-\sum\limits_{i=1}^L \frac{(\Delta+1)\alpha_i}{2}} \prod_{i<j} (\rho^{-\alpha_i}-\rho^{-\alpha_j}) d \ba \\
&\geq \frac{(1-\rho^{-\delta})^{\frac{L(L-1)}{2}}}{e^L}  \int_{\mathcal{A}_0 \cap S_{\delta}}  \rho^{-\sum\limits_{i=1}^L\alpha_i N_i} d\ba \doteq  \int_{\mathcal{A}_0 \cap S_{\delta}} \rho^{-\sum\limits_{i=1}^L\alpha_i N_i} d\ba,
\end{align*} 
}%
where $N_i=\frac{1}{2}(\Delta+2L-2i+1)$. The previous inequality follows from the fact that $\rho^{-\alpha_i}-\rho^{-\alpha_j}>\rho^{-\alpha_i}(1-\rho^{-\delta})$ for $\ba \in S_{\delta}$, and $e^{-\rho^{-\alpha_i}}\geq \frac{1}{e}$ if $\alpha_i \geq 0$. (Note that for a fixed $i$, there are $L-i$ possible values for $j$ such that $i<j$.)

\begin{lemma} \label{inf_lemma}
Let $f(\ba)=\sum\limits_{i=1}^L (q+L+1-2i)\alpha_i$. 
Then 
$$\inf\limits_{\ba \in \mathcal{A}_0}  f(\ba)=(-q-L+2\floor{s}+1)s+qL-\floor{s}(\floor{s}+1)=f(\boldsymbol\alpha^*),$$
where $\alpha_1^*=\ldots=\alpha_{k-1}^*=0$, $\alpha_k^*=k-s$, $\alpha_{k+1}^*=\ldots=\alpha_L^*=1$ for $k=\floor{s}+1$.
\end{lemma}

The proof of Lemma \ref{inf_lemma} 
can be found in Appendix \ref{proof_inf_lemma}.\\
Using Lemma \ref{inf_lemma} with $q=\Delta+L$, $s=2r$, we find that $\inf_{\ba \in \mathcal{A}_0} \sum_{i=1}^L N_i \alpha_i=\inf_{\ba \in \mathcal{A}_0} \frac{f(\ba)}{2}$ is equal to
{\allowdisplaybreaks
\begin{align*}
&\frac{1}{2}\left[(-\Delta-2L+2\floor{2r}+1)2r+(\Delta+L)L-\floor{2r}(\floor{2r}+1)\right]\\
&=(-2m-n+2\floor{2r}+1)r+mn-\frac{\floor{2r}(\floor{2r}+1)}{2}.
\end{align*}
}%
This is the piecewise function $d_1(r)$ connecting the points $(r,[(m-r)(n-2r)]^+)$ where $2r \in \Z$.\\
Using the Laplace principle, $\forall \delta>0$ we have
$$\lim_{\rho \to \infty} -\frac{\log P_{\out}(R)}{\log\rho}\geq \inf_{\mathcal{A}_0 \cap S_{\delta}} \frac{f(\boldsymbol\alpha)}{2}.$$
Note that $\forall \delta$, the point
$\ba_{\delta}$ such that $\alpha_{\delta,i}=\alpha_i^*+\frac{\delta i}{L}$ is in $\mathcal{A}_0 \cap S_{\frac{\delta}{L}}$ and when $\delta \to 0$, $\ba_{\delta} \to \ba^*$. By continuity of $f$,
\begin{align}
\lim_{\delta \to 0} \inf_{\mathcal{A}_0 \cap S_{\delta}} \frac{f(\ba)}{2}=\frac{f(\ba^*)}{2}=d_1(r). \tag*{\IEEEQED}
\end{align}
\let\IEEEQED \relax%
\end{IEEEproof}

\subsection{DMT of real lattice codes with NVD} \label{real_lower_bound}
In this section, we show that real spherically shaped lattice codes with the NVD property achieve the DMT upper bound of Theorem \ref{theorem_real_upper}. This result extends Proposition 4.2 in \cite{ISIT2016_MIMO}.
\begin{thm} \label{theorem_real_lower}
Let $\mathcal{L}$ be an $n^2$-dimensional lattice in $M_n(\R)$, and consider the 
code $\mathcal{C}(\rho)=\rho^{-\frac{r}{n}} \mathcal{L}(\rho^{\frac{r}{n}})$. If $\mathcal{L}$ has the NVD property, then the DMT of the code $\mathcal{C}(\rho)$ is 
the function $d_1(r)$ connecting the points $(r,[(m-r)(n-2r)]^+)$ where $2r \in \Z$.
\end{thm}

\begin{IEEEproof}
Since the upper bound has already been established in Theorem \ref{theorem_real_upper}, we only need to prove that the DMT is lower bounded by $d_1(r)$. The following section follows very closely the proof in \cite{EKPKL}, and thus some details are omitted. To simplify notation, we assume that $\mindet{\mathcal{L}}=1$.\\
We consider the sphere bound for the error probability for the equivalent real channel (\ref{real_channel}): for a fixed channel realization $H$,
$$P_e(H) \leq \mathbb{P}\left\{ \norm{W}^2 > d_H^2/4\right\}$$
where $d_H^2$ is the 
squared 
minimum distance in the received constellation: 
\begin{align*}
&d_H^2=\frac{\rho}{n} \min_{\bar{X},\bar{X}' \in \mathcal{C}(\rho),\;\bar{X} \neq \bar{X}'} \norm{H(\bar{X}-\bar{X}')}^2
=\frac{1}{n}\rho^{1-\frac{2r}{n}} \min_{X,X' \in \mathcal{L}(\rho^{\frac{r}{n}}),\;X \neq X'} \norm{H(X-X')}^2.  
\end{align*}
We denote $\Delta X=X-X'$. Let $L=\min(2m,n)$, and $\Delta=\abs{n-2m}$. Let $\lambda_1 \geq \lambda_2 \geq \cdots \geq \lambda_L >0$ be the nonzero eigenvalues of $H^T H$, and $0 \leq \mu_1 \leq \cdots \leq \mu_n$ 
the eigenvalues of $\Delta X \Delta X^T$. Using the mismatched eigenvalue bound and the arithmetic-geometric inequality as in \cite{EKPKL}, 
for all $k=1, \ldots, L$
{\allowdisplaybreaks
\begin{align*}
&d_H^2=\frac{1}{n}\rho^{1-\frac{2r}{n}}\min_{X,X' \in \mathcal{L}(\rho^{\frac{r}{n}}),\;X \neq X'} \tr(H\Delta X \Delta X^T H^T)
\geq \frac{1}{n}\rho^{1-\frac{2r}{n}} \sum_{i=1}^L \mu_i \lambda_i \geq \frac{k}{n} \rho^{1-\frac{2r}{n}}  \left(\prod_{i=1}^k \lambda_i\right)^{\frac{1}{k}} \left(\prod_{i=1}^k \mu_i\right)^{\frac{1}{k}}.
\end{align*} 
}
For all $i= 1, \ldots, n$, $\mu_i \leq \norm{\Delta X}^2 \leq 
4 \rho^{\frac{2r}{n}}$, and 
$$ \prod_{i=1}^n \mu_i =\det(\Delta X \Delta X^T) \geq 1$$
due to the NVD property. Consequently, for all $k=1,\ldots,L$
$$\prod_{i=1}^k \mu_i =\frac{\det(\Delta X \Delta X^T)}{\prod_{i=k+1}^n \mu_i} \geq \frac{1}{4^{n-k}\rho^{\frac{2r(n-k)}{n}}}.$$ 
With the change of variables $\lambda_i=\rho^{-\alpha_i}$ $\forall i=1,\ldots,L$, we can write 
{\allowdisplaybreaks
\begin{equation} \label{eq_dH}
d_H^2 \geq \frac{k}{n 4^{\frac{n-k}{k}}}  \rho^{1-\frac{2r}{n}} \rho^{-\frac{1}{k} \sum\limits_{i=1}^k \alpha_i} \frac{1}{\rho^{\frac{2r(n-k)}{nk}}}= c_k \rho^{-\frac{1}{k}\left(\sum\limits_{i=1}^k \alpha_i +2r -k \right)}
= c_k\rho^{\delta_k(\boldsymbol\alpha,s)} \quad \forall k=1,\ldots,L,
\end{equation}
}
where we have set $\boldsymbol\alpha=(\alpha_1,\ldots,\alpha_L)$, $s=2r$, $c_k=\frac{k}{n 4^{\frac{n-k}{k}}}$ and 
\begin{equation} \label{delta}
\delta_k(\boldsymbol\alpha,s)=-\frac{1}{k}\left(\sum\limits_{i=1}^k \alpha_i +s -k \right).
\end{equation}
Since $2 \norm{W}^2$ is a $\chi^2(2mn)$ random variable, we have
$$\mathbb{P}\left\{\norm{W}^2 >d\right\}=\Phi_{mn}(d),$$
where we define
\begin{equation} \label{Phi}
\Phi_t(d)=\sum_{i=0}^{t-1} e^{-d} \frac{d^i}{i!}.
\end{equation}
Let $p(\ba)$ be the distribution of $\ba$ in (\ref{p_alpha_real}). For $i<j$, $\rho^{-\alpha_i} \geq \rho^{-\alpha_j}$ and for a fixed $i$, there are $L-i$ possible values for $j$, and 
\begin{equation} \label{p_prime}
p(\boldsymbol\alpha) \leq p'(\boldsymbol\alpha)=K e^{-\sum\limits_{i=1}^L \rho^{-\alpha_i}} \rho^{-\sum\limits_{i=1}^L\alpha_i N_i}(\log \rho)^L
\end{equation}
where $N_i=\frac{1}{2}(\Delta+2L-2i+1)$. By averaging over the channel, the error probability is bounded by
\begin{equation} \label{17b}
P_e =\int_{\mathcal{A}} P_e(\boldsymbol\alpha) p(\boldsymbol\alpha) d\boldsymbol\alpha \leq \int_{\mathcal{A}} \mathbb{P}\left\{ \norm{W}^2 > \frac{d_H^2}{4}\right\} p(\boldsymbol\alpha) d\boldsymbol\alpha \leq \int_{\mathcal{A}} \Phi_{mn}\left(\frac{d_H^2}{4}\right) p'(\boldsymbol\alpha) d\boldsymbol\alpha
\end{equation}
where $\mathcal{A}=\{\ba\;:\; \alpha_1 \leq \cdots \leq \alpha_L\}$.

The following Lemma closely follows \cite{EKPKLold}, which is a preliminary version of \cite{EKPKL}, and it is proven in Appendix \ref{proof_laplace_lemma}:

\begin{lemma} \label{laplace_lemma}
Assuming that $d \geq c_k\rho^{\delta_k(\boldsymbol\alpha,s)}$ for some constants $c_k$, $k=1,\ldots,L$, then for all $t \in \N^+$,  
{\allowdisplaybreaks
\begin{align*}
&-\lim_{\rho \to \infty} \frac{1}{\log \rho} \log \int_{\mathcal{A}} p'(\ba) \Phi_t\left(d\right) d\ba 
\geq \inf_{\ba \in \mathcal{A}_0} \sum_{i=1}^L N_i \alpha_i,
\end{align*}
}
where $\mathcal{A}_0$ is defined in (\ref{A_0}). 
\end{lemma}

The proof of the Theorem is concluded using Lemma \ref{inf_lemma} with $q = \Delta+L$, $s = 2r$.
\end{IEEEproof}
\begin{figure}[tb]
\begin{scriptsize}
\begin{center}
\begin{tikzpicture}[xscale=3.8,yscale=0.6]
\draw[->] (0,0) -- (2.1,0);
\draw (2.1,0) node[right] {$r$};
\draw [->] (0,0) -- (0,8.9);
\draw (0,0) node[below] {$0$};
\draw (0.5,0) node[below] {$\frac{1}{2}$};
\draw (1,0) node[below] {$1$};
\draw (0,0.5) node[left] {$\frac{1}{2}$};
\draw (0,2) node[left] {$2$};
\draw (0,3) node[left] {$3$};
\draw (0,4.5) node[left] {$\frac{9}{2}$};
\draw (1.5,0) node[below] {$\frac{3}{2}$};
\draw (2,0) node[below] {$2$};
\draw (0,8) node[left] {$8$};
\draw (0,8.9) node[above] {$d(r)$};
\draw [dotted] (0.5,0) -- (0.5,4.5);
\draw [dotted] (1,0) -- (1,3);
\draw [dotted] (0,0.5) -- (1.5,0.5) ;
\draw [dotted] (1.5,0) -- (1.5,0.5);
\draw [dotted] (0,2) -- (1,2);
\draw [dotted] (0,3) -- (1,3);
\draw [dotted] (0,4.5) -- (0.5,4.5);
\draw [thick, samples=200, domain=0:2] plot(\x,{(-7+2*floor(2*\x))*\x+8-floor(2*\x)*(floor(2*\x)+1)/2});
\draw [dashed, thick, samples=200, domain=0:2] plot(\x,{2*((-1)*(3-2*floor(\x))*\x+4-floor(\x)*(floor(\x)+1))});
\draw [dotted, thick, samples=200, domain=0:2] plot(\x,{(-5+2*floor(\x))*\x+8-floor(\x)*(floor(\x)+1)});
\end{tikzpicture}
\end{center}
\end{scriptsize}
\caption{DMT upper bounds for real (solid) and quaternion (dashed) codes for $n=4$ and $m=2$. The dotted lines correspond to the 
optimal DMT.} 
\end{figure}
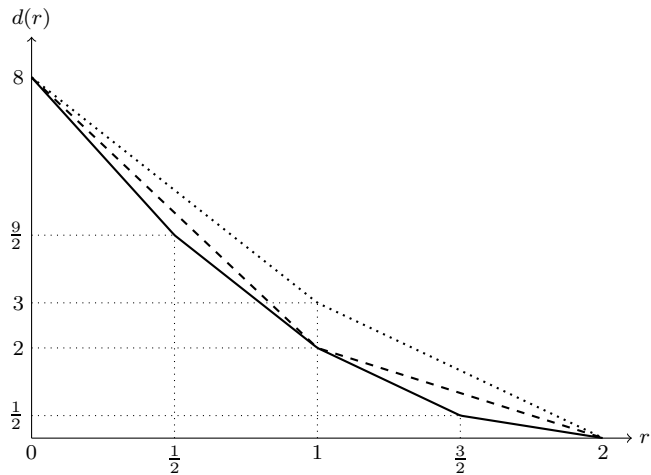

\subsection{Equivalent channel model for quaternion lattice codes}
Suppose that $n=2p$ is even. We consider again the channel 
\begin{equation} \label{channel_2}
Y_c=\sqrt{\frac{\rho}{n}} H_c \bar{X} + W_c,
\end{equation}
and we suppose that the codewords $\bar{X}$ are of the form
$$\bar{X}=\begin{pmatrix} A & -B^* \\ B & A^* \end{pmatrix} \in M_{2p}(\C),$$
where $A,B \in M_p(\C)$.

First, we derive an equivalent model where the channel has quaternionic form. We can write 
$$Y_c=\begin{pmatrix} Y_1 & Y_2\end{pmatrix}, \quad H_c=\begin{pmatrix} H_1 & H_2\end{pmatrix}, \quad W_c=\begin{pmatrix} W_1 & W_2\end{pmatrix},$$
where $Y_1, Y_2, H_1, H_2, W_1, W_2 \in M_{m \times p}(\C)$. Then 
$$Y_1\!=\!\sqrt{\frac{\rho}{n}}(H_1 A + H_2 B) + W_1, \; Y_2\!=\!\sqrt{\frac{\rho}{n}}(-H_1 B^* + H_2 A^*)+W_2,$$
and we have the equivalent \vv{quaternionic channel}:
$$\underbrace{\begin{pmatrix} Y_1 & Y_2 \\ -Y_2^* & Y_1^* \end{pmatrix}}_{\text{\normalsize{$Y$}}} =\sqrt{\frac{\rho}{n}} \underbrace{\begin{pmatrix} H_1 & H_2 \\ -H_2^* & H_1^* \end{pmatrix}}_{\text{\normalsize{$H$}}} \underbrace{\begin{pmatrix} A & -B^* \\ B & A^* \end{pmatrix}}_{\text{\normalsize{$\bar{X}$}}} + \underbrace{\begin{pmatrix} W_1 & W_2 \\ -W_2^* & W_1^* \end{pmatrix}}_{\text{\normalsize{$W$}}}.$$

\subsection{General DMT upper bound for quaternion codes}

\begin{thm} \label{theorem_quaternion_upper}
Suppose that $\forall \rho$, $\mathcal{C}(\rho) \subset M_{n/2}(\H)$. Then the DMT of the code $\mathcal{C}$ is upper bounded by the function $d_2(r)$ connecting the points $(r,[(m-r)(n-2r)]^+)$ for $r \in \Z$.
\end{thm}

\begin{IEEEproof} The quaternionic channel can be written in the complex MIMO channel form
\begin{equation} \label{quaternion_channel_2}
\begin{pmatrix} Y_1 \\ -Y_2^* \end{pmatrix} =\sqrt{\frac{\rho}{n}} \begin{pmatrix} H_1 & H_2 \\ -H_2^* & H_1^* \end{pmatrix} \begin{pmatrix} A \\ B \end{pmatrix} + \begin{pmatrix} W_1 \\ -W_2^* \end{pmatrix}
\end{equation}
If $r$ is the multiplexing gain of the original system (\ref{channel_2}), then the multiplexing gain of this channel is $2r$, since the same number of symbols is transmitted using half the frame length.\\
Consider the eigenvalues  $\lambda_1=\lambda_1' \geq \lambda_2=\lambda_2' \geq \cdots \geq \lambda_p=\lambda_p' \geq 0$ of $H^{\dagger}H$. Let $L=\min(m,p)$ the number of pairs of nonzero  eigenvalues, and $\Delta=\abs{p-m}$.
For fixed $H$, the capacity of this channel is \cite{Telatar}
$$C(H) \doteq \log \det (I+ \rho H^{\dagger} H)=2 \sum_{i=1}^L \log (1+\rho \lambda_i ).$$
The joint eigenvalue density $p(\boldsymbol\lambda)=p(\lambda_1,\ldots,\lambda_L)$ of a quaternion Wishart matrix is \cite{Edelman_Rao}\footnote{The quaternion case corresponds to taking $\beta=4$ in \cite[equation (4.5)]{Edelman_Rao}. Note that we modify the distribution to take into account the fact that each entry of $H$ has variance $1/2$ per real dimension.}
$$p(\lambda_1,\ldots,\lambda_L)=K \prod_{i<j} (\lambda_i -\lambda_j)^4 \prod_{i=1}^L \lambda_i^{2\Delta+1}e^{-\sum\limits_{i=1}^L \lambda_i}$$  
for some constant $K$. With the change of variables $\lambda_i=\rho^{-\alpha_i}$ $\forall i=1,\ldots,L$, the distribution of $\ba=(\alpha_1,\ldots,\alpha_L)$ is
$$p(\ba)\!\!=\!\!K(\log \rho)^le^{-\sum\limits_{i=1}^L \rho^{-\alpha_i}}\! \rho^{-2\sum\limits_{i=1}^L \alpha_i (\Delta+1)}\!\prod_{i<j}\!\left(\rho^{-\alpha_i}\!\!-\!\!\rho^{-\alpha_j}\right)^4$$%
The outage probability for rate $R=r \log \rho$ is given by
{\allowdisplaybreaks
\begin{align*}
&P_{\out}(R)\doteq \mathbb{P}\left\{2 \sum_{i=1}^L \log(1+\rho \lambda_i)<2r \log \rho\right\}
=\!\mathbb{P}\left\{\prod_{i=1}^L (1\!+\!\rho^{1-\alpha_i})\!<\!\rho^{r}\!\right\}\!\doteq\! \mathbb{P}\left\{\prod_{i=1}^L \rho^{(1-\alpha_i)^+}\!\!<\!\rho^{r}\!\right\} \!\geq\! \mathbb{P}(\mathcal{A}_0)
\end{align*}
}%
where $\mathcal{A}_0=\left\{\ba: 0 \leq \alpha_1 \leq \ldots \leq \alpha_L,\; \sum\limits_{i=1}^L (1-\alpha_i)^+ <r\right\}$.
Given $\delta >0$, define $S_{\delta}=\{\ba:\; \abs{\alpha_i-\alpha_j}> \delta \; \forall i \neq j\}$. Then
{\allowdisplaybreaks
\begin{align*}
&P_{\out}(R)
\dotgeq \!\!\int_{\mathcal{A}_0 \cap S_{\delta}}\! e^{-\sum\limits_{i=1}^L \rho^{-\alpha_i}} \!\rho^{-2\sum\limits_{i=1}^L\alpha_i(\Delta +1)} \prod_{i<j} (\rho^{-\alpha_i} -\rho^{-\alpha_j})^4 d\ba
\geq \frac{(1-\rho^{-\delta})^{\frac{L(L-1)}{2}}}{e^{L}} \int_{\mathcal{A}_0 \cap S_{\delta}} \rho^{-\sum\limits_{i=1}^L N_i \alpha_i} d\ba
\end{align*}}%
where $N_i=2(\Delta+2L-2i+1)$.
Let $f(\ba)=\sum_{i=1}^L (q+L-2i+1)$. Using the Laplace principle, 
$$\lim\limits_{\rho \to \infty} -\frac{\log P_{\out}(R)}{\log\rho}\geq 2\inf_{\mathcal{A}_0 \cap S_{\delta}} f(\alpha) \; \forall \delta>0.$$
Using Lemma \ref{inf_lemma} with $s=r$, $q=\Delta+L$, we find that $\inf_{\ba \in \mathcal{A}_0} N_i \alpha_i=2\inf_{\ba \in \mathcal{A}_0} f(\ba)=2f(\ba^*)$ is the piecewise linear function $d_2(r)$ connecting the points 
$(r, \left[2(p-r)(m-r)\right]^+)=(r,\left[(n-2r)(m-r)\right]^+)$ for $r \in \Z$. Note that $\forall \delta$, the point
$\ba_{\delta}$ such that $\ba_{\delta,i}=\alpha_i^*+\frac{\delta i}{L}$ is in $\mathcal{A}_0 \cap S_{\frac{\delta}{L}}$ and when $\delta \to 0$, $\ba_{\delta} \to \ba^*$. By continuity of $f$, $2\lim_{\delta \to 0} \inf_{\mathcal{A}_0 \cap S_{\delta}} f(\ba)=2f(\ba^*)=d_2(r)$.
\end{IEEEproof}

\subsection{DMT of quaternionic lattice codes with NVD}
 
We now show that quaternionic lattice codes with NVD achieve the upper bound of Theorem \ref{theorem_quaternion_upper}.  This result extends Proposition 4.3 in \cite{ISIT2016_MIMO}. 
 
\begin{thm} \label{theorem_quaternion_lower}
Let $\mathcal{L}$ be an $n^2$-dimensional lattice in $M_{n/2}(\H)$ with the NVD property.
Then the DMT of the code $\mathcal{C}(\rho)=\rho^{-\frac{r}{n}} \mathcal{L}(\rho^{\frac{r}{n}})$
is 
the piecewise linear function $d_2(r)$ connecting the points $(r,[(m-r)(n-2r)]^+)$ for $r \in \Z$.
\end{thm}

\begin{IEEEproof}
To simplify notation, assume $\mindet{\mathcal{L}}=1$. 
For a fixed 
realization $H$, $P_e(H) \leq \mathbb{P}\left\{ \norm{W}^2 > d_H^2/4\right\}$,
where 
{\allowdisplaybreaks
\begin{align*}
&d_H^2 =\frac{1}{n}\rho^{1-\frac{2r}{n}} \min_{X,X' \in \mathcal{L}(\rho^{\frac{r}{n}}),\;X \neq X'} \norm{H(X-X')}^2.
\end{align*}
}%
Let $\Delta X=X-X'$. 
We denote by $\lambda_1=\lambda_1' \geq \lambda_2=\lambda_2' \geq \cdots \geq \lambda_p=\lambda_p' \geq 0$  the eigenvalues of $H^{\dagger} H$, and by $0 \leq \mu_1=\mu_1' \leq \cdots \leq \mu_p=\mu_p'$ the eigenvalues of $\Delta X \Delta X^{\dagger}$. Both sets of eigenvalues have multiplicity $2$ since $H$ and $X$ are quaternion matrices. Again we set $L=\min(m,p)$ and $\Delta=\abs{p-m}$. Using the mismatched eigenvalue bound and the arithmetic-geometric inequality as in \cite{EKPKL}, 
for all $k=1,\ldots,L$,
{\allowdisplaybreaks
\begin{align*}
&d_H^2=\frac{1}{n}\rho^{1-\frac{2r}{n}} \min_{X,X' \in \mathcal{C}(\rho),\;X \neq X'} \tr(H\Delta X \Delta X^{\dagger} H^{\dagger}) 
\geq \frac{1}{n}\rho^{1-\frac{2r}{n}} \sum_{i=1}^L (2 \mu_i \lambda_i) \geq \frac{k}{p} \rho^{1-\frac{2r}{n}} \left(\prod_{i=1}^k \lambda_i\right)^{\frac{1}{k}} \left(\prod_{i=1}^k \mu_i\right)^{\frac{1}{k}}. 
\end{align*}
}%
For all $i=1,\ldots,p$, $\mu_i \leq \norm{\Delta X}^2 \leq 4 \rho^{\frac{2r}{n}}$, and 
$ \prod_{i=1}^p \mu_i =\det(\Delta X \Delta X^{\dagger})^{\frac{1}{2}} \geq 1$
using the NVD property of the code. For all $k=1,\ldots,L$
$$\prod_{i=1}^k \mu_i =\frac{\det(\Delta X \Delta X^{\dagger})^{\frac{1}{2}}}{\prod_{i=k+1}^p \mu_i} \geq \frac{1}{4^{p-k}\rho^{\frac{2r(p-k)}{n}}}=\frac{1}{4^{p-k}\rho^{\frac{r(p-k)}{p}}}.$$ 
With the change of variables $\lambda_i=\rho^{-\alpha_i}$ $\forall i=1,\ldots,L$, we have 
\begin{equation}
d_H^2 \geq c_k\rho^{1-\frac{r}{p}} \rho^{-\frac{1}{k} \sum\limits_{i=1}^k \alpha_i} \rho^{-\frac{r(p-k)}{pk}}\!= c_k\rho^{-\frac{1}{k}\big(\sum\limits_{i=1}^k \alpha_i +r -k \big)}\!\!=c_k \rho^{\delta_k(\boldsymbol\alpha)} \quad\forall k=1,\ldots,L, \label{eq_dH_2}
\end{equation}
where $\boldsymbol\alpha=(\alpha_1,\ldots,\alpha_L)$, $\delta_k(\boldsymbol\alpha,r)=-\frac{1}{k}\left(\sum\limits_{i=1}^k \alpha_i +r -k \right)$, and $c_k=\frac{k}{p 4^{\frac{p-k}{k}}}$.\\
Since $2 \norm{W}^2 \sim 2 \chi^2(2mp)$, we have
{\allowdisplaybreaks
\begin{align*}
&P_e(H) \leq \mathbb{P}\left\{ \norm{W}^2 > \frac{d_H^2}{2}\right\}
=\Phi_{mp}\left(\frac{d_H^2}{2}\right),
\end{align*}
}
where $\Phi_t$ is the function defined in (\ref{Phi}).
By averaging with respect to the distribution $p(\ba)$, we get $\forall k=1,\ldots,L$
\begin{equation*} 
P_e \leq \int_{\mathcal{A}} p(\boldsymbol\alpha) \Phi_{mp}\left(\frac{d_H^2}{2}\right)d\boldsymbol\alpha \dotleq  \int_{\mathcal{A}} p'(\boldsymbol\alpha) \Phi_{mp}\left(\frac{d_H^2}{2}\right)d\boldsymbol\alpha
\end{equation*}
where $\mathcal{A}=\{\ba: \alpha_1 \leq \cdots \leq \alpha_L\}$, and
$$p'(\boldsymbol\alpha)=K(\log \rho)^L e^{-\sum\limits_{i=1}^L \rho^{-\alpha_i}}\ \rho^{-\sum\limits_{i=1}^L \alpha_i N_i},$$
where $N_i=2(\Delta+2L-2i+1)$. Note that $p'(\ba)$ has the same form as (\ref{p_prime}), and the condition (\ref{eq_dH_2}) is of the same form as (\ref{eq_dH}). From Lemma \ref{laplace_lemma} we find $d(r) \geq \inf_{\ba \in \mathcal{A}_0}  2\sum_{i=1}^L \alpha_i (\Delta+2L-2i+1)$,
which by Lemma \ref{inf_lemma} is the piecewise linear function connecting the points 
$(r,[(n-2r)(m-r)]^+)$ for $r \in \Z$.
\end{IEEEproof}

\section{Division Algebra Codes Achieve the Optimal  Restricted DMT in $M_{n/2}(\H)$ and $M_n(\R)$}\label{division_algebra}

 Theorems  \ref{theorem_real_lower} and \ref{theorem_quaternion_lower} state that $n^2$-dimensional NVD lattices  in  $M_n(\R)$ and $M_{n/2}(\H)$ do achieve the respective DMT upper bounds of Theorems \ref{theorem_real_upper} and \ref{theorem_quaternion_upper}. In order to show that these bounds are tight and indeed describe the optimal restricted DMTs, it is enough to prove the existence of $n^2$-dimensional  NVD lattice codes in $M_n(\R)$ and $M_{n/2}(\H)$. For that we need some results from non-commutative algebra. For details and definitions  we refer the reader to \cite{Reiner}.

Let  $\mathcal{D}$ be an index $n$   $\Q$-central division algebra. We say that $\D$ is \emph{ramified at the infinite place} if 
$$
\D\otimes_{\Q}\R\simeq M_{n/2}(\mathbb{H}).
$$
If it is not, then
$$
\D\otimes_{\Q}\R\simeq M_{n}(\R).
$$

Let $\Lambda$ be an \emph{order} in an index $n$ $\Q$-central division algebra $\mathcal{D}$. We then have the following.
\begin{lemma} \cite{VHO}\label{embeddings}
If the infinite prime is ramified in the algebra $\D$, then there exists an embedding
$$\psi_{\mathrm{abs}}: \mathcal{D} \to M_{n/2}(\mathbb{H})$$
 such that $\psi_{\mathrm{abs}}(\Lambda)$ is an $n^2$-dimensional NVD lattice.
If $\D$ is not ramified at the infinite place, then there exists an embedding
$$\psi_{\mathrm{abs}}: \mathcal{D} \to M_{n}(\R)$$
such that $\psi_{\mathrm{abs}}(\Lambda)$ is an $n^2$-dimensional NVD lattice.
For every $n$ there exists an index $n$   $\Q$-central  division algebra that is ramified at the infinite place and one which is not.
\end{lemma}

The following corollary follows from Theorems  \ref{theorem_real_lower} and    \ref{theorem_quaternion_lower} and from Lemma \ref{embeddings}. It proves that the upper bounds in Theorems \ref{theorem_real_upper} and \ref{theorem_quaternion_upper} are tight.

\begin{corollary}\label{corollary_quaternion}
For every $n$ there exists an $n^2$-dimensional  NVD lattice $\mathcal{L}\subset M_n(\R)$ that  achieves the upper bound of Theorem \ref{theorem_real_upper}. For every even  $n$ there exists an $n^2$-dimensional  NVD lattice $\mathcal{L}\subset M_{n/2}(\H)$ that  achieves the upper bound of Theorem \ref{theorem_quaternion_upper}.
\end{corollary}

The following corollary gives us a complete DMT characterization of $\Q$-central division algebra codes. The DMT of such codes only depends on whether the  corresponding algebra is ramified at the infinite place or not.

\begin{corollary}\label{corollary_division_both}
Let $\Lambda$ be an order in an index $n$ $\Q$-central division algebra $\mathcal{D}$.
If $\D$ is ramified at the infinite place, then the code $\psi_{\mathrm{abs}}(\Lambda)\subset M_{n/2}(\H)$     achieves the upper bound of Theorem \ref{theorem_quaternion_upper}.
If $\D$ is not ramified at the  infinite place, then the DMT of the code  $\psi_{\mathrm{abs}}(\Lambda)\subset M_n(\R)$  achieves the upper bound of  Theorem \ref{theorem_real_upper}.   
\end{corollary}


\subsection{DMT of $\Q$ central division algebra codes based on the regular representation}\label{algebra_DMT}
In the previous sections we classified the DMT of all $\Q$-central division algebra codes. However, this result was proven in the case where the code lattices were  constructed  using the abstract embedding of  Lemma \ref{embeddings}. In contrast, explicit codes are typically built using \emph{regular representations}. In this section we study the DMT of division algebra codes that are constructed by using such representations.

Let  $E/\Q$ be a cyclic field extension of degree $n$ with  Galois group $ G(E/\Q)= \langle\sigma\rangle $. Define a cyclic algebra
$$
\D=(E/\Q,\sigma,\gamma)=E\oplus uE\oplus u^2E\oplus\cdots\oplus u^{n-1}E,
$$
where   $u\in\mathcal{D}$ is an auxiliary
generating element subject to the relations
$xu=u\sigma(x)$ for all $x\in E$ and $u^n=\gamma\in \Q^*$. 

Considering $\D$ as a right  vector space over $E$, every element $x=x_0+ux_1+\cdots+u^{n-1}x_{n-1}\in\mathcal{D}$
has the following left regular representation as a matrix $\psi_{reg}(x)$:
\begin{equation*}\label{presentation}
\begin{pmatrix}
x_0& \gamma\sigma(x_{n-1})& \gamma\sigma^2(x_{n-2})&\cdots &
\gamma\sigma^{n-1}(x_1)\\
x_1&\sigma(x_0)&\gamma\sigma^2(x_{n-1})& &\gamma\sigma^{n-1}(x_2)\\
x_2& \sigma(x_1)&\sigma^2(x_0)& &\gamma\sigma^{n-1}(x_3)\\
\vdots& & & & \vdots\\
x_{n-1}& \sigma(x_{n-2})&\sigma^2(x_{n-3})&\cdots&\sigma^{n-1}(x_0)\\
\end{pmatrix}.
\end{equation*}

The  mapping $\psi_{reg}$ is an injective $\Q$-algebra homomorphism  that allows us to identify
$\D$ with its image in $M_n(\C)$.

\begin{proposition}\cite{Reiner}\label{Zlattice}
If $\Lambda$ is a  $\Z$-order in an index $n$ $\Q$-central division algebra $\D$, then 
$\psi_{reg}(\Lambda)$ is  an $n^2$-dimensional NVD lattice in $M_n(\C)$.
 \end{proposition}

\begin{exam}
Consider the following two algebras
$$
\A_1=(\Q(\sqrt{3})/\Q, \sigma, -1)\,\,\mathrm{and}\,\,\A_2=(\Q(i)/\Q, \sigma, -1).
$$
Let us use the notation $\Z[\sqrt{3}]=\Z+\Z\sqrt{3}$ and  $\Z[i]=\Z+i\Z$. By using regular presentation $\psi_{reg}$, we can  find the following $4$-dimensional lattice  codes
\begin{align*}
&\mathcal{L}_1=\left\{\begin{pmatrix}
x_1&-x_2\\
x_2& x_1
\end{pmatrix}:\quad x_1,x_2 \in \Z[\sqrt{3}]\right\},\\
&\mathcal{L}_2=\left\{\begin{pmatrix}
x_1&-x_2^*\\
x_2& x_1^*
\end{pmatrix}:\quad x_1,x_2 \in \Z[i]\right\}.
\end{align*} 
Here $\mathcal{L}_1$  corresponds to the algebra $\A_1$ \cite{BelReal}, while $\mathcal{L}_2$ corresponds to  the algebra $\A_2$ and is the lattice of the Alamouti code. As $\mathcal{L}_1$ is completely real and $\mathcal{L}_2$ is quaternionic, we can read their DMTs from Theorems \ref{theorem_real_lower} and \ref{theorem_quaternion_lower}. Here the DMT of Alamouti was already known, while the DMT of $\mathcal{L}_2$ is a new result.
\end{exam}

However, in general, while the lattices of Proposition \ref{Zlattice} have the correct dimension and the NVD property, there is no guarantee that they are always contained in $M_n(\R)$ or in $M_{n/2}(\H)$ and we can not directly apply Theorems \ref{theorem_real_lower} and \ref{theorem_quaternion_lower}. However, the following result shows that all the lattices produced by regular representations are conjugated versions of lattices whose DMT we know:

\begin{lemma} \cite[Lemma 9.10]{VLL2013}\label{embeddings2}
 Let $\D$ be an index $n$ $\Q$-central division algebra  and $\Lambda \subset \D$ an order. If the infinite prime is ramified in the algebra $\D$, then there exists an invertible matrix $A\in M_n(\C)$ such that
$$
A\psi_{reg}(\Lambda)A^{-1}=\psi_{abs}(\Lambda)\subset M_{n/2}(\mathbb{H}).
$$
If $\D$ is not ramified at the infinite place, then there exists an invertible  matrix $B\in M_n(\C)$ such that
$$
B\psi_{reg}(\Lambda)B^{-1}=\psi_{abs}(\Lambda)\subset M_{n}(\R).
$$
\end{lemma}
The following  conjecture then seems to be plausible, but its proof has eluded us.

\begin{conjecture}\label{conjecture:conjugate}
Let $\D$ be an index $n$ $\Q$-central division algebra  and $\Lambda \subset \D$ an order. If $\D$ is ramified  at the infinite prime, then $\psi_{reg}(\Lambda)$ achieves the DMT upper bound of Theorem \ref{theorem_quaternion_upper}. If $\D$ is not ramified at the infinite prime, then $\psi_{reg}(\Lambda)$ achieves the DMT of Theorem \ref{theorem_real_upper}.
\end{conjecture}

\begin{exam}
Applying the regular representation  to the algebra $\D_1=(\Q(i)/\Q, \sigma, 3)$ yields the following lattice
\begin{equation*}
\mathcal{L}_1=\left\{\begin{pmatrix}
x_1&3x_2^*\\
x_2& x_1^*
\end{pmatrix}:\quad x_1,x_2 \in \Z[i]\right\}.
\end{equation*}
We can easily see that  $\D_1$ is not ramified at the infinite place, but on the other hand   $\mathcal{L}_1 \nsubseteq M_2(\R)$. However, our conjecture claims that the DMT of $\mathcal{L}_1$ is described by Theorem  \ref{theorem_real_upper}.
\end{exam}


\section{Multi-block codes}\label{multi-block}

When introducing the concept of diversity-multiplexing in \cite{ZT} the authors mostly focused on one shot quasi-static channels. However, they also considered a channel model where it is possible to decode and encode over  a fixed number of independent faded blocks and found the corresponding optimal DMT curve.

In this section we consider such multi-block channels
\begin{equation} \label{multi-block_channel}
Y_c^{(l)}=\sqrt{\frac{\rho}{n}} H_c^{(l)} \bar{X}^{(l)} + W_c^{(l)}, \quad l=1,\ldots,k,
\end{equation}
where $H_c^{(l)}, W_c^{(l)} \in M_{m,n}(\C)$ are the channel and noise matrices with i.i.d. circularly symmetric complex Gaussian entries in $\mathcal{N}_{\C}(0,1)$. The set of multi-block codewords $X=[X^{(1)},\ldots,X^{(k)}]$ should satisfy the global power constraint
\begin{equation} \label{multi-block_power_constraint}
\frac{1}{kn^2} \frac{1}{\abs{\mathcal{C}}} \sum_{X \in \mathcal{C}} \sum_{l=1}^k \bnorm{X^{(l)}}_F^2 \leq 1. 
\end{equation}
A {\em multi-block matrix lattice} $\mathcal{L} \subseteq M_{n\times nk}(\C)$ has the form
$$\mathcal{L}=\Z B_1\oplus \Z B_2\oplus \cdots \oplus \Z B_d,$$
where the matrices $B_1,\dots, B_d \in M_{n \times nk}(\C)$ are linearly independent over $\R$, and $d \leq 2 n^2 k$ is the dimension of the lattice. 

We then have a natural extension for the NVD condition.  First we define 
$$
\mathrm{pdet}(X)=\prod_{i=1}^{k} \mathrm{det}(X^i).
$$

\begin{definition}\label{pmindet}
Given a multi-block lattice $\mathcal{L} \subseteq M_{n\times nk}(\C)$,  we say that the lattice satisfies the \emph{non-vanishing determinant} (NVD) property if 
\[
\inf_{X \in L \setminus \{\bf 0\}} \abs{\mathrm{pdet}(X)}>0.
\]
\end{definition}

Given a multi-block lattice $\mathcal{L}\subseteq M_{n\times nk}(\C)$ of dimension $d$, we consider spherically shaped multi-block codes of the form
\begin{equation} \label{multi-block_spherical_shaping}
\mathcal{C}(\rho)=\rho^{-\frac{rnk}{d}}\mathcal{L}(\rho^{\frac{rnk}{d}}).
\end{equation} 
Note that such a code will satisfy the power constraint (\ref{multi-block_power_constraint}), and its multiplexing gain per block is $r$.

A general DMT upper bound for multi-block codes $\mathcal{C} \subset M_n(\C)^k$ was given in \cite[Section V]{ZT}. In \cite{Lu2008} it was proven that  $2n^2k$-dimensional lattice multi-block codes with the NVD property achieve this DMT upper bound, extending the result of \cite{EKPKL} to the multi-block case. However, as in the case of the single block channel, the DMT of asymmetric multi-block codes is mostly unknown.
  
We will now consider  multi-block codes that are subsets of  $M_{n\times n}(\R)^k$ or  $M_{n/2}(\H)^k$, and show that if the codewords of a space-time code belong to either of these spaces, its DMT is limited by a bound that is tighter than the general DMT bound and depends on the ambient space. We then show that if  a space-time lattice code belongs to $M_{n}(\R)^k$ or  $M_{n/2}(\H)^k$, has degree $n^2k$ and satisfies the NVD condition, it achieves the corresponding restricted DMT. Furthermore, we prove that division algebra based codes do achieve these restricted DMT limits for every $k$ and $n$.



 Let us now assume we have a degree $k$ number field  $K$  with signature $(r_1, r_2)$, and   an index $n$ $K$-central division algebra $\D$.
 We then have that
\begin{equation}\label{formal}
\D\otimes_\Q\R \cong M_{n/2}(\H)^{\omega} \times M_{n}(\R)^{r_1 -\omega} \times M_n(\C)^{2r_2},
\end{equation}
where $\omega\leq r_1$ is an integer depending on the structure of the algebra $\D$. 
We call the triplet $(\omega, r_1-\omega, r_2)$  the \emph{signature} of the algebra  $\D$. We note that this result is an extension of Lemma \ref{embeddings2}. The signature of  $\Q$ is $(1,0)$. Hence any $\Q$-central division algebra has signature $(\omega, 1-\omega,0)$.  When $\omega=1$ the algebra is ramified at the infinite prime and when $\omega=0$ it is not.

\begin{proposition}\cite{VHO}\label{psiabs}
Let $\D$ be a $K$-central division algebra with signature $(\omega, r_1-\omega, r_2)$ of index $n$ and $\Lambda$  an order in $\D$. Then $\psi_{abs}(\Lambda)$
is a $kn^2$ dimensional lattice in $ M_{n/2}(\H)^{\omega} \times M_{n}(\R)^{r_1 -\omega} \times M_n(\C)^{2r_2}$ and
$$
\mindet{\psi_{abs}(\Lambda)}= 1.
$$
\end{proposition}

\begin{lemma}\label{multi-block_existence}
For   any integer $n$ and triplet $(\omega, r_1-\omega, r_2)$ there exist a number field $K$  and  a $K$-central index $n$ division algebra $\D$ with signature $(\omega, r_1-\omega, r_2)$.
\end{lemma}
In particular, according to Proposition \ref{psiabs}, for any $n$ (respectively for any even $n$) and for any $k$, there exists a $kn^2$-dimensional multi-block code with NVD in $M_{n}(\R)^k$ (respectively in $M_{n/2}(\H)^{k}$).

\subsection{Real multi-block codes}

We have the following multi-block extensions of Theorems \ref{theorem_real_upper} and \ref{theorem_real_lower}:

\begin{thm} \label{theorem_real_upper_multi-block}
Suppose that $\forall \rho$, $\mathcal{C}(\rho) \subset M_n(\R)^{k}$. Then the DMT of the code $\mathcal{C}$ is upper bounded by $kd_1(r)$, where $d_1(r)$ is the function connecting the points $(r,[(m-r)(n-2r)]^+)$ for $2r \in \Z$.
\end{thm}

\begin{thm} \label{theorem_real_lower_multi-block}
Let $\mathcal{L}$ be an $n^2k$-dimensional lattice in $M_n(\R)^k$, and consider the 
spherically shaped 
code $\mathcal{C}(\rho)=\rho^{-\frac{r}{n}} \mathcal{L}(\rho^{\frac{r}{n}})$. If $\mathcal{L}$ has the NVD property, then the DMT of the code $\mathcal{C}(\rho)$ is 
the function $k d_1(r)$.
\end{thm}

The proof of Theorems \ref{theorem_real_upper_multi-block} and \ref{theorem_real_lower_multi-block} can be found in Appendix \ref{proof_multi-block_real}. 

We then have the following corollary that follows directly from Theorem \ref{theorem_real_lower_multi-block} and Lemma \ref{multi-block_existence}.

\begin{corollary}
For every $n$ and $k$ there exists a $kn^2$-dimensional  NVD lattice $\mathcal{L}\subset M_{n}(\R)^k$ that  achieves the DMT of Theorem \ref{theorem_real_upper_multi-block}.
\end{corollary}

\subsection{Quaternion multi-block codes}

Similarly, we can extend Theorems \ref{theorem_quaternion_upper} and \ref{theorem_quaternion_lower} to the multi-block case:

\begin{thm} \label{theorem_quaternion_upper_multi-block}
Suppose that $\forall \rho$, $\mathcal{C}(\rho) \subset M_{n/2}(\H)^{k}$. Then the DMT of the code $\mathcal{C}$ is upper bounded by $kd_2(r)$, where $d_2(r)$ is the function connecting the points $(r,[(m-r)(n-2r)]^+)$ for $r \in \Z$.
\end{thm}

\begin{thm} \label{theorem_quaternion_lower_multi-block}
Let $\mathcal{L}$ be an $n^2k$-dimensional lattice in $M_{n/2}(\H)^k$, and consider the 
spherically shaped 
code $\mathcal{C}(\rho)=\rho^{-\frac{r}{n}} \mathcal{L}(\rho^{\frac{r}{n}})$. If $\mathcal{L}$ has the NVD property, then the DMT of the code $\mathcal{C}(\rho)$ is 
the function $k d_2(r)$.
\end{thm}

The proof of these Theorems can be found in Appendix \ref{proof_multi-block_quaternion}. 

According to  Lemma  \ref{multi-block_existence} we now have the following.
\begin{corollary}
For every even $n$ and any $k$ there exists a $kn^2$-dimensional  NVD lattice $\mathcal{L}\subset M_{n/2}(\H)^k$ that  achieves the DMT of Theorem \ref{theorem_quaternion_upper_multi-block}.
\end{corollary}

\appendix

\subsection{Proof of Lemma \ref{inf_lemma}} \label{proof_inf_lemma}
Let $\bar{d}(s)=(-q-L+2\floor{s}+1)s+qL-\floor{s}(\floor{s}+1)$. Without loss of generality, we can suppose that $k-1 \leq s < k$ for some $k \in \N$, i.e. $k-1=\floor{s}$, $k=\floor{s}+1$.\\
First, we show that $\forall \ba \in \mathcal{A}_0$, we have $f(\ba) \geq \bar{d}(s)$. In fact
{\allowdisplaybreaks
\begin{align*}
&f(\ba)=\left(q-L-1\right)\sum\limits_{i=1}^L \alpha_i +2\sum\limits_{i=1}^L (L-i+1)\alpha_i  
=\left(q-L-1\right)\sum\limits_{i=1}^L \alpha_i +2\sum\limits_{i=1}^L \sum_{j=1}^i\alpha_j \\
& \geq \left(q-L-1\right)(L-s)+2\sum\limits_{i=k}^L \sum_{j=1}^i\alpha_j 
\geq \left(q-L-1\right)(L-s)+2\sum\limits_{i=k}^L (i-s)\\
&=\left(q-L-1\right)(L-s)+L(L+1)-(k-1)k-2(L-k+1)s
=\bar{d}(s).
\end{align*} 
}
Next, we show that $\exists \ba^*$ such that $f(\ba^*)=\bar{d}(s)$.\\ 
Let $\alpha_1^*=\ldots=\alpha_{k-1}^*=0$, $\alpha_k^*=k-s$, $\alpha_{k+1}^*=\ldots=\alpha_L^*=1$. Then
{\allowdisplaybreaks
\begin{align*}
&f(\ba^*)=\sum_{i=1}^L \left(q+L+1\right)\alpha_i-2\sum_{i=1}^L i \alpha_i 
= \left(q+L+1\right)(k-s)+ \left(q+L+1\right)(L-k) -2k(k-s)-2\!\!\sum_{i=k+1}^L i \\
&=\left(q+L+1\right)(L-s)-2k(k-s) -L(L+1)+k(k+1)
=\bar{d}(s) \tag*{\IEEEQED}
\end{align*}
}

\subsection{Proof of Lemma \ref{laplace_lemma}} \label{proof_laplace_lemma} 
The proof closely follows \cite{EKPKLold}, which is a preliminary version of \cite{EKPKL}. Note that $\Phi_t\left(d\right) \leq 1$ since it is a probability. Given $\varepsilon>0$, we can bound the integral (\ref{17b}) as follows
\begin{equation} \label{integral_2}
\int_{\mathcal{A}} p'(\boldsymbol\alpha) \Phi_t\left(d\right)d\boldsymbol\alpha \leq \int_{\bar{\mathcal{A}}} p'(\boldsymbol\alpha) \Phi_t\left(d\right)d\boldsymbol\alpha + \sum_{j=1}^L \int_{\mathcal{A}_j} p'(\boldsymbol\alpha) \Phi_t\left(d\right)d\boldsymbol\alpha,
\end{equation}
where $\bar{\mathcal{A}}=\{\ba \in \mathcal{A}\;:\;\alpha_i \geq -\varepsilon \;\;\forall i=1,\ldots,L\}$ and $\mathcal{A}_j=\{\ba \in \mathcal{A}\;:\;\alpha_j < - \varepsilon\}$. 
Note that
{\allowdisplaybreaks
\begin{align*}
&\int_{\mathcal{A}_j} p'(\boldsymbol\alpha) \Phi_t\left(d\right)d\boldsymbol\alpha \leq \int_{\mathcal{A}_j} p'(\boldsymbol\alpha)d\boldsymbol\alpha 
\dotleq \left(\prod_{i \neq j} \int_{-\infty}^{\infty} e^{-\rho^{-\alpha_i}} \rho^{-\alpha_i N_i} d\alpha_i\right)\int_{-\infty}^{-\varepsilon} e^{-\rho^{-\alpha_j}}\rho^{-\alpha_j N_j} d\alpha_j\\
&=\left(\prod_{i \neq j} \int_{0}^{\infty} \frac{e^{-\lambda_i} \lambda_i^{N_i-1}}{\log \rho}d\lambda_i\right)\int_{\rho^{\varepsilon}}^{\infty} \frac{\lambda_j^{N_j-1} e^{-\lambda_j}}{\log \rho} d\lambda_j 
\doteq \rho^{0} \int_{\rho^{\varepsilon}}^{\infty} \frac{\lambda_j^{N_j-1} e^{-\lambda_j}}{\log \rho} d\lambda_j 
\end{align*}
}
which vanishes exponentially fast as a function of $\rho$. For the first term in (\ref{integral_2}), we have
{\allowdisplaybreaks
\begin{equation*}
\int_{\bar{\mathcal{A}}} p'(\boldsymbol\alpha) \Phi_t\left(d\right)d\boldsymbol\alpha \leq \int\limits_{\substack{\ba > -\varepsilon \\ \boldsymbol\delta(\alpha,s)< \varepsilon}} p'(\boldsymbol\alpha) \Phi_t\left(d\right) d\ba 
+\sum_{j=1}^L \int\limits_{\substack{\ba > -\epsilon,\\ \delta_j(\ba,s)\geq \varepsilon}}p'(\boldsymbol\alpha) \Phi_t\left(d\right) d\ba,
\end{equation*}
}%
where the notation $\ba > -\epsilon$ means $\alpha_i >-\epsilon \;\;\forall i=1,\ldots,L$, and $\boldsymbol\delta(\boldsymbol\alpha,s)=(\delta_1(\boldsymbol\alpha,s),\ldots,\delta_L(\boldsymbol\alpha,s))$. Since $\Phi_t(d)$ is a decreasing function of $d$, using the assumption that $d \geq c_j\rho^{\delta_j(\boldsymbol\alpha,s)}\; \forall j=1,\ldots,L$, (\ref{eq_dH}) we can write
{\allowdisplaybreaks
\begin{align}
&\int\limits_{\substack{\ba > -\epsilon,\\ \delta_j(\ba,s)\geq \varepsilon}}p'(\boldsymbol\alpha) \Phi_t\left(d\right) d\ba
\dotleq \int\limits_{\substack{\ba > -\epsilon,\\ \delta_j(\ba,s)\geq \varepsilon}}p'(\boldsymbol\alpha) \Phi_t\left(c_j \rho^{\delta_j(\boldsymbol\alpha,s)}\right) d\ba \notag\\
&\dotleq \left(\prod_{i=j+1}^{L} \int\limits_{\alpha_i >-\varepsilon} \rho^{-\alpha_i N_i} d\alpha_i\right) 
\cdot\!\! \int\limits_{\substack{\alpha_1,\ldots,\alpha_j >- \varepsilon\\ \delta_j(\ba,s) \geq \varepsilon}}\!\!e^{-c_j \rho^{\delta_j(\boldsymbol\alpha,s)}}\sum_{\tau=0}^{t-1}\!\!\left(c_j \rho^{\delta_j(\boldsymbol\alpha,s)}\right)^{\tau} \frac{1}{\tau!} \rho^{-\sum\limits_{i=1}^j \alpha_i N_i}d\alpha_1 \ldots d\alpha_j \label{d}
\end{align}
}%
since $\delta_j(\ba,s)$ is independent of $\alpha_i$ for $i >j$. As $\delta_j(\ba,s) \geq \varepsilon$, $\alpha_i>-\varepsilon$,
 the second integral is over a bounded region and tends to zero exponentially fast as a function of $\rho$, while the first integral has a finite SNR exponent. Thus, (\ref{d}) tends to zero exponentially fast. \\
Finally, the SNR exponent of (\ref{17b}) is determined by the behavior of 
{\allowdisplaybreaks 
\begin{align*}
&\int\limits_{\substack{\ba > -\varepsilon \\ \boldsymbol\delta(\ba,s)< \varepsilon}} p'(\boldsymbol\alpha) \Phi_t\left(d\right) d\ba \leq \int\limits_{\substack{\ba > -\varepsilon \\ \boldsymbol\delta(\ba,s)< \varepsilon}} p'(\boldsymbol\alpha) d\ba 
\dotleq \int\limits_{\substack{\ba > -\varepsilon \\ \boldsymbol\delta(\ba,s)< \varepsilon}} \rho^{-\sum\limits_{i=1}^n N_i \alpha_i} d\ba.
\end{align*}
}
The conclusion follows by using the Laplace principle, and taking $\epsilon \to 0$. Note that 
{\allowdisplaybreaks
\begin{align}  
&\mathcal{A}_0=\left\{\ba \in \mathcal{A}:\;\alpha_j\geq 0,\;\sum_{i=1}^j (1-\alpha_i) \leq s \; \forall j=1,\ldots,L\right\} \notag\\
&=\{\ba: \alpha_j \geq 0, \; \delta_j(\ba,s) \leq 0 \;\; \forall j=1,\ldots,L\}. \tag*{\IEEEQED}
\end{align}
}

\subsection{Proof of Theorems \ref{theorem_real_upper_multi-block} and \ref{theorem_real_lower_multi-block} (DMT of real multi-block codes)} \label{proof_multi-block_real}
Consider a multi-block lattice $\mathcal{L} \subset M_n(\R)^k$ of dimension $d=n^2 k$, and a multi-block code $\mathcal{C}(\rho)=\rho^{-\frac{r}{n}} \mathcal{L}(\rho^{\frac{r}{n}})$. Every codeword is of the form $X=[X^{(1)},\ldots,X^{(k)}]$. \\ 
Similarly to the single-block case, for all $l=1,\ldots,k$ we can write 
$$Y_c^{(l)}=Y_{r}^{(l)} + i Y_{i}^{(l)}, \quad H_c^{(l)}=H_{r}^{(l)} + i H_{i}^{(l)}, \quad W_c^{(l)}=W_{r}^{(l)} + i W_{i}^{(l)}$$
and obtain the equivalent real channel with $2m$ receive antennas:
$$Y_c^{(l)}=\begin{pmatrix} Y_{r}^{(l)} \\ Y_{i}^{(l)} \end{pmatrix}=\sqrt{\frac{\rho}{n}} \begin{pmatrix} H_{r}^{(l)} \\ H_{i}^{(l)} \end{pmatrix} X^{(l)} + \begin{pmatrix} W_{r}^{(l)} \\ W_{i}^{(l)} \end{pmatrix}= H^{(l)} X^{(l)}+ W^{(l)},$$
where $H^{(l)} \in M_{2m \times n}(\R)$, $W^{(l)} \in M_{2m \times n}(\R)$ have real i.i.d. Gaussian entries with variance $1/2$.

\subsubsection{Proof of Theorem \ref{theorem_real_upper_multi-block}}
We can write the outage probability as 
$$P_{\out}(R)=\mathbb{P}\left\{ \frac{1}{k} \left(\frac{1}{2}\sum_{l=1}^{k} \log\det(I+\rho (H^{(l)})^{T}H^{(l)})\right) \leq R\right\}.$$
Define $L=\min(2m,n)$, $\Delta=\abs{n-2m}$, and let  
\begin{align*}
& \lambda_1^{(l)} \geq \cdots \geq \lambda_{L}^{(l)},\quad l=1,\ldots,k
\end{align*}
the ordered nonzero eigenvalues of $(H^{(l)})^{T}H^{(l)}$. Their distribution is
\begin{align*}
&p(\lambda_1^{(l)},\ldots,\lambda_{L}^{(l)})= K \prod_{i=1}^{L} (\lambda_i^{(l)})^{\frac{\Delta-1}{2}} e^{-\sum\limits_{i=1}^{L} \lambda_i^{(l)}} \prod_{i < j} \abs{\lambda_i^{(l)} - \lambda_{j}^{(l)}}, \quad l=1,\ldots,k.
\end{align*}
Thus, we have 
$$P_{\out}(R)=\mathbb{P}\left\{ \prod_{l=1}^{k} \prod_{i=1}^{L} (1+\rho \lambda_i^{(l)})^{1/2}  \leq \rho^{rk}\right\}.$$
Consider the change of variables $\lambda_i^{(l)}=\rho^{-\alpha_i^{(l)}} \; \forall l=1,\ldots,k$, and let
$$\mathcal{A}=\left\{ \boldsymbol{\alpha} \in \R^{kL} \; :\; 
 0 \leq \alpha_1^{(l)} \leq \cdots \leq \alpha_{L}^{(l)} \;\; \forall l=1,\ldots,k
\right\}$$
Then
\begin{equation} \label{product_distribution}
p(\boldsymbol\alpha) \doteq \rho^{-\sum\limits_{l=1}^k\sum\limits_{i=1}^L \frac{\Delta+1}{2}\alpha_i^{(l)}}e^{-\sum\limits_{l=1}^k \sum\limits_{i=1}^L \rho^{-\alpha_i^{(l)}}} \prod_{l=1}^k \prod_{i<j} \abs{\rho^{-\alpha_i^{(l)}}-\rho^{-\alpha_{j}^{(l)}}}.
\end{equation}
Recalling that $1+\rho^{1-x} \dotleq \rho^{(1-x)^+}$, we have
\begin{align*}
&P_{\out}(R)=\mathbb{P}\left\{ \prod_{l=1}^{k} \prod_{i=1}^{L} (1+\rho^{1-\alpha_i^{(l)}})^{1/2} \leq \rho^{rk}\right\}\geq \mathbb{P}(\mathcal{A}_0),
\end{align*}
where 
\begin{align*}
& \mathcal{A}_0=\left\{ \boldsymbol\alpha \in \mathcal{A} \;:\; \frac{1}{2} \sum_{l=1}^{k} \sum_{i=1}^{L} (1-\alpha_i^{(l)})^+ \leq rk\right\}\\
&=\left\{ \boldsymbol\alpha \in \mathcal{A} \;:\; \forall \mathbf{j}=(j_1,\ldots,j_k) \,:\, j_l \leq L \;\;\forall l, \; \frac{1}{2} \sum_{j=1}^{k} \sum_{i=1}^{j_l} (1-\alpha_i^{(l)}) \leq rk\right\} 
\end{align*}
Given $\delta >0$, let
$ \mathcal{S}_{\delta}=\left\{ \boldsymbol\alpha \in \mathcal{A} \;:\; \forall i \neq j, \;
\abs{\alpha_i^{(l)} - \alpha_{j}^{(l)}} > \delta \;\;  \forall l=1,\ldots,k \right\}$.
Then
\begin{align*}
&P_{\out}(R) \dotgeq \int_{\mathcal{A}_0} p(\boldsymbol\alpha) d\boldsymbol\alpha \geq \int_{\mathcal{A}_0 \cap S_{\delta}} p(\boldsymbol\alpha) d\boldsymbol\alpha  \\
&\doteq\int_{\mathcal{A}_0 \cap S_{\delta}} \rho^{-\sum\limits_{l=1}^{k} \sum\limits_{i=1}^{L} \alpha_i^{(l)} \frac{\Delta+1}{2} } e^{-\sum\limits_{l=1}^{k} \sum\limits_{i=1}^{L} \rho^{-\alpha_i^{(l)}}} 
\prod_{l=1}^{k} \prod_{1 \leq i<j \leq L} \abs{\rho^{-\alpha_i^{(l)}}-\rho^{-\alpha_{j}^{(l)}}} d\boldsymbol\alpha  \\
&\dotgeq \frac{(1-\rho^{-\delta})^{k\frac{L(L-1)}{2}}}{e^{Lk}}\int_{\mathcal{A}_0 \cap S_{\delta}} \rho^{-\frac{1}{2} \sum\limits_{l=1}^{k} \sum\limits_{i=1}^{L} (\Delta + 2 L - 2i +1) \alpha_i^{(l)}}d\boldsymbol\alpha.
\end{align*}     
To find the DMT upper bound, we need an extension of Lemma \ref{inf_lemma} to the multi-block case:
\begin{lemma} \label{inf_lemma_multi-block}
Let $F(\boldsymbol\alpha)=\sum_{l=1}^k \sum_{i=1}^L (q+L -2 i + 1) \alpha_i^{(l)}$.\\
Then $\inf_{\boldsymbol\alpha \in \mathcal{A}_0} F(\boldsymbol\alpha)=k\left[(-q-L+2 \floor{s} +1)s+qL-\floor{s}(\floor{s}+1)\right]=k \bar{d}(s).$
\end{lemma}
\begin{IEEEproof}[Proof of Lemma \ref{inf_lemma_multi-block}]
Note that if
\begin{align*}
\bar{\alpha}_1^{(l)}=\cdots=\bar{\alpha}_{\floor{s}}^{(l)}=0, \quad \bar{\alpha}_{\floor{s}+1}^{(l)}=1+\floor{s}-s,\quad \bar{\alpha}_{\floor{s}+2}^{(l)}=\cdots=\bar{\alpha}_{L}^{(l)}=1 \quad \forall l=1,\ldots,k,
\end{align*}
then $\bar{\boldsymbol\alpha} \in \mathcal{A}_0$ and $F(\bar{\boldsymbol\alpha})=k\left[(-q-L+2 \floor{s} +1)s+qL-\floor{s}(\floor{s}+1)\right]$. We want to show that this value is the minimum of the function $F$ over $\mathcal{A}_0$. \\
Note that for $\boldsymbol\alpha \in \mathcal{A}_0$ we have the following global constraints: $\forall j \leq L$,
\begin{equation}
\sum_{l=1}^{k}  \sum_{i=1}^{j}\alpha_i^{(l)} \geq k(j-s). \label{global_constraints}
\end{equation}
Recalling that 
$ \sum\limits_{i=1}^{L} \sum\limits_{j=1}^i \alpha_j=\sum\limits_{i=1}^L (L-i+1) \alpha_i,$
we can write
\begin{align*}
&F(\boldsymbol\alpha)=(q-L-1) \sum_{l=1}^{k} \sum_{i=1}^{L} \alpha_i^{(l)} +2 \sum_{l=1}^k \sum_{i=1}^L (L-i+1) \alpha_i^{(l)} 
=(q-L -1) \sum_{l=1}^k \sum_{i=1}^{L} \alpha_i^{(l)} + 2\sum_{l=1}^{k} \sum_{i=1}^{L} \sum_{j=1}^i \alpha_j^{(l)}\\
&\geq k(q-L -1) (L-s) + 2 \sum_{i=1}^L k(i-s)=k \bar{d}(s),
\end{align*}
where the final step in the proof is the same as in Lemma \ref{inf_lemma}.
\end{IEEEproof}
Using Lemma \ref{inf_lemma_multi-block} with $q=\Delta +L$, $s=2r$, we find that the DMT upper bound $\inf_{\boldsymbol\alpha \in \mathcal{A}_0} \frac{F(\boldsymbol\alpha)}{2}=k d_1(r)$. This concludes the proof of Theorem \ref{theorem_real_upper_multi-block}. \hfill \IEEEQED

\subsubsection{Proof of Theorem \ref{theorem_real_lower_multi-block}}
The proof for the lower bound is similar to the proof of Theorem 2 in \cite{Lu2008}, but we include it for completeness\footnote{Note that compared to \cite{Lu2008}, we deal separately with the eigenvalues in each block instead of re-ordering them. The two approaches are equivalent.}. Letting $H=\diag(H^{(1)},\ldots,H^{(k)})$ and  $W=[W^{(1)},\ldots,W^{(k)}]$ the multi-block channel matrix and noise for the equivalent real channel, we have the sphere bound $P_e(H) \leq \mathbb{P}\{ \norm{W}^2 > d_H^2/4\}$,
where 
$$d_H^2=\frac{\rho}{n} \min_{\substack{X, X' \in \mathcal{C}(\rho) \\ X \neq X'}} \sum_{l=1}^k \norm{H^{(l)}(X^{(l)}-X'^{(l)})}^2 \geq \frac{1}{n}\rho^{1-\frac{2r}{n}} \sum_{l=1}^k \sum_{i=1}^L \lambda_i^{(l)} \mu_i^{(l)},$$
where 
$0 \leq \mu_1^{(l)} \leq \cdots \leq \mu_n^{(l)}$ are the ordered eigenvalues of $\Delta X^{(l)} (\Delta X ^{(l)})^T$ with $\Delta X=X-X'$. \\
For any $\mathbf{j}=(j_1,\ldots,j_k)$ with $J=\sum_{l=1}^k j_l \geq 1$, we have
\begin{align*}
d_H^2 \geq \frac{1}{n} \rho^{1-\frac{2r}{n}} \sum_{l=1}^k \sum_{i=1}^{j_l} \lambda_i^{(l)} \mu_i^{(l)} \geq \frac{1}{n}\rho^{1-\frac{2r}{n}} \frac{J}{n}\left( \prod_{l=1}^k \prod_{i=1}^{j_l} \lambda_i^{(l)} \mu_i^{(l)}\right)^{\frac{1}{J}}.
\end{align*}
Note that $\forall i=1,\ldots, L$, $\forall l=1,\ldots,k$, $\mu_i^{(l)} \leq 4 \rho^{\frac{2r}{n}}$, and
$$\prod_{l=1}^k \prod_{i=1}^{j_l} \mu_i^{(l)} =\frac{\det(\Delta X \Delta X^T)}{\prod_{l=1}^k \prod_{i=j_l+1}^n \mu_i^{(l)}} \geq \frac{1}{4^{kn-J}\rho^{(kn-J)\frac{2r}{n}}}.$$
Therefore $\forall \mathbf{j} \neq 0$,
\begin{align} \label{eq_dH_multiblock}
d_H^2 \geq c_j \rho^{\delta_{\mathbf{j}}(\boldsymbol\alpha,2r)}
\end{align}
where $\delta_{\mathbf{j}}(\boldsymbol\alpha,s)=-\frac{1}{\sum_{l=1}^k j_l}\sum\limits_{l=1}^k \left(\sum\limits_{i=1}^{j_l} \alpha_i^{(l)} + s -j_l\right)$, $s=2r$, and $c_j$ is a suitable constant. \\
The proof proceeds similarly to Section \ref{real_lower_bound}. 
We have
$$P_e =\int P_e(\boldsymbol\alpha) p(\boldsymbol\alpha) d\boldsymbol\alpha \leq \int \mathbb{P} \left\{ \norm{W}^2 > \frac{d_H^2}{4}\right\} p(\boldsymbol \alpha) d \boldsymbol\alpha.$$
Note that the distribution $p(\boldsymbol\alpha)$ in (\ref{product_distribution}) is upper bounded as follows: 
$$p(\boldsymbol\alpha) \dotleq p'(\boldsymbol\alpha)=e^{-\sum\limits_{l=1}^k \sum\limits_{i=1}^L \rho^{-\alpha_i^{(l)}}} \rho^{-\sum\limits_{l=1}^k \sum\limits_{i=1}^L \alpha_i^{(l)} N_i}$$
where $N_i=\frac{1}{2}(\Delta+ 2L - 2i +1)$. \\
Since $2\norm{W}^2 \sim \chi^2(2mnk)$, we have $\mathbb{P}\left\{ \norm{W}^2 > \frac{d_H^2}{4}\right\} =\Phi_{mnk}\left(\frac{d_H^2}{4}\right)$, where $\Phi_t$ is defined in (\ref{Phi}). So $\forall \mathbf{j} \neq \mathbf{0}$,
$$P_e \leq \int_{\mathcal{A}} p'(\boldsymbol\alpha) \Phi\left(\frac{d_H^2}{4}\right) d \boldsymbol\alpha.$$
To conclude the proof, we need an extension of Lemma \ref{laplace_lemma} to the multi-block case:
\begin{lemma} \label{laplace_lemma_multi-block}
Assuming that $d \geq c_j \rho^{\delta_{\mathbf{j}}(\boldsymbol\alpha,s)} \; \forall \mathbf{j} \neq \mathbf{0}$
, then $\forall t \in \mathbb{N}^+$,
$$ -\lim_{\rho \to \infty} \frac{1}{\log \rho} \log \int_{\mathcal{A}} p'(\boldsymbol\alpha) \Phi\left(d\right) d\boldsymbol\alpha \geq \inf_{\boldsymbol\alpha \in \mathcal{A}_0} \sum\limits_{l=1}^k \sum\limits_{i=1}^L N_i \alpha_i^{(l)},$$
where $\mathcal{A}_0=\left\{ \boldsymbol\alpha \in \mathcal{A} \;:\; \forall \mathbf{j}=(j_1,\ldots,j_k) \,:\forall l,\, j_l \leq L, \;  \sum\limits_{l=1}^{k} \sum\limits_{i=1}^{j_l} (1-\alpha_i^{(l)}) \leq sk\right\}$. 
\end{lemma}
\begin{IEEEproof}[Proof of Lemma \ref{laplace_lemma_multi-block}]
The proof is very similar to the proof of Lemma \ref{laplace_lemma}. We include a sketch for convenience. \\
Note that $\Phi_t\left(\frac{d_H^2}{4}\right) \leq 1$ since it is a probability. If we define
\begin{align*}
&\bar{\mathcal{A}}=\left\{ \boldsymbol\alpha \in \mathcal{A} \;:\; \alpha_i^{(l)} \geq - \epsilon \; \forall i=1,\ldots,L, \; \forall k=1,\ldots,l\right\},\\ 
& \mathcal{A}_i^{(l)}=\left\{ \boldsymbol\alpha \in \mathcal{A} \; : \; \alpha_i^{(l)} < -\epsilon\right\},
\end{align*}
then we have the bound
\begin{equation} \label{star_multi-block}
P_e \leq \int_{\bar{\mathcal{A}}} p'(\boldsymbol\alpha) \Phi_t\left(d\right) d\boldsymbol\alpha + 
\sum\limits_{l=1}^l \sum\limits_{i=1}^L \int_{\mathcal{A}_i^{(l)}} p'(\boldsymbol\alpha) \Phi_t\left(d\right)d \boldsymbol\alpha
\end{equation}
With the change of variables $\lambda_i^{(l)}=\rho^{-\alpha_i^{(l)}}$ \, $\forall l=1,\ldots,k$, $\forall i=1,\ldots,L$, we have
\begin{align*}
&\int_{\mathcal{A}_i^{(l)}} p'(\boldsymbol\alpha) \Phi_t\left(d\right) d\boldsymbol\alpha \leq \int_{\mathcal{A}_i^{(l)}} p'(\boldsymbol\alpha) d\boldsymbol\alpha \dotleq 
\Bigg(\prod_{(i',l') \neq (i,l)} \int_{-\infty}^{\infty} e^{-\rho^{-\alpha_{i'}^{(l')}}} \rho^{-\alpha_{i'}^{(l')} N_{i'}} d\alpha_{i'}^{(l')} \Bigg) \int_{-\infty}^{-\epsilon} e^{-\rho^{-\alpha_{i}^{(l)}}} \rho^{-\alpha_{i}^{(l)} N_{i}} d\alpha_{i}^{(l)}=\\
& \Bigg(\prod_{(i',l') \neq (i,l)} \int_{0}^{\infty} \frac{e^{-\lambda_{i'}^{(l')}} (\lambda_{i'}^{(l')})^{N_{i'}-1}d\lambda_{i'}^{(l')}}{\log \rho}\Bigg) \left(\int_{\rho^{\epsilon}}^{\infty} \frac{e^{-\lambda_i^{(l)}} (\lambda_i^{(l)})^{N_i-1} d\lambda_i^{(l)}}{\log \rho}\right) \doteq \int_{\rho^{\epsilon}}^{\infty} \frac{e^{-\lambda_i^{(l)}} (\lambda_i^{(l)})^{N_i-1} d\lambda_i^{(l)}}{\log \rho}
\end{align*}
which vanishes exponentially as a function of $\rho$. The first term in (\ref{star_multi-block}) is bounded by
\begin{align}  \label{square_multi-block}
&\int_{\bar{\mathcal{A}}} p'(\boldsymbol\alpha) \Phi_t\left(d\right) d\boldsymbol\alpha \leq  
\int_{\substack{\boldsymbol\alpha >-\epsilon,\\ \delta_{\mathbf{j}'}(\boldsymbol\alpha,s) < \epsilon \; \forall \mathbf{j'}}} p'(\boldsymbol\alpha) \Phi_t\left(d\right) d\boldsymbol\alpha + \int_{\substack{\boldsymbol\alpha >-\epsilon,\\ \delta_{\mathbf{j}'}(\boldsymbol\alpha,s) \geq \epsilon \; \forall \mathbf{j'}}} p'(\boldsymbol\alpha) \Phi_t\left(d\right) d\boldsymbol\alpha. 
\end{align}
Since $\Phi_t$ is decreasing, and using the assumption that $d \geq c_j \rho^{\delta_{\mathbf{j}}(\boldsymbol\alpha,s)} \; \forall \mathbf{j} \neq \mathbf{0}$, we have 
\begin{align*}
& \int_{\substack{\boldsymbol\alpha >-\epsilon,\\ \delta_{\mathbf{j}'}(\boldsymbol\alpha,s) \geq \epsilon \; \forall \mathbf{j'}}} p'(\boldsymbol\alpha) \Phi_t\left(d\right) d\boldsymbol\alpha \leq \int_{\substack{\boldsymbol\alpha >-\epsilon,\\ \delta_{\mathbf{j}'}(\boldsymbol\alpha,s) \geq \epsilon \; \forall \mathbf{j'}}} p'(\boldsymbol\alpha) \Phi_t\left(c_j\rho^{\delta_{\mathbf{j}}(\boldsymbol\alpha,s)}\right) d\boldsymbol\alpha  \\
& \dotleq \int_{\substack{\boldsymbol\alpha >-\epsilon,\\ \delta_{\mathbf{j}'}(\boldsymbol\alpha,s) \geq \epsilon \; \forall \mathbf{j'}}} e^{-c_j\rho^{\delta_{\mathbf{j}}(\boldsymbol\alpha,s)}} \sum_{\tau=0}^{t-1} \left(c_j \rho^{\delta_{\mathbf{j}}(\boldsymbol\alpha,s)}\right)^{\tau} \frac{1}{\tau!} \prod_{l=1}^k \prod_{i=1}^L \rho^{-\alpha_i^{(l)} N_i} d\boldsymbol\alpha \\
& \dotleq \Bigg( \prod_{l=1}^k \prod_{i >j_l} \int_{\alpha_i^{(l)}>-\epsilon} \rho^{-\alpha_i^{(l)} N_i} d\alpha_i^{(l)}\Bigg) 
\Bigg(\int_{\substack{\alpha_i^{(l)}>-\epsilon \;\; \forall i<j_l\\ \delta_{\mathbf{j}'}(\boldsymbol\alpha,s) \geq \epsilon}} e^{-c_j \rho^{\delta_{\mathbf{j}}(\boldsymbol\alpha,s)}} \sum_{\tau=0}^{t-1} \left(c_j\rho^{\delta_{\mathbf{j}}(\boldsymbol\alpha,s)}\right)^{\tau} \frac{1}{\tau!} \rho^{-\sum\limits_{l=1}^k \sum\limits_{i=1}^{j_l} N_i} \prod_{l=1}^k \prod_{i=1}^{j_l} d\alpha_i^{(l)}
\Bigg)
\end{align*}
since $\delta_{\mathbf{j}'}(\boldsymbol\alpha,s)$ is independent of $\alpha_i^{(l)}$ $\forall i> j_l'$.
The first integral has a finite SNR exponent, while the second is over a bounded region, and so it tends to $0$ exponentially as a function of $\rho$. Thus, the product also tends to zero exponentially.\\
To conclude, observe that the first term in (\ref{square_multi-block}) is upper bounded by
\begin{align*}
\int_{\substack{\boldsymbol\alpha >-\epsilon,\\ \delta_{\mathbf{j}'}(\boldsymbol\alpha,s) < \epsilon \; \forall \mathbf{j'}}} p'(\boldsymbol\alpha) d\boldsymbol\alpha \dotleq \int_{\substack{\boldsymbol\alpha >-\epsilon,\\ \delta_{\mathbf{j}'}(\boldsymbol\alpha,s) < \epsilon \; \forall \mathbf{j'}}} \rho^{-\sum\limits_{l=1}^k \sum\limits_{i=1}^L \alpha_i^{(l)} N_i} d \boldsymbol\alpha.
\end{align*}
The statement follows by using the Laplace principle and taking $\epsilon \to 0$. 
\end{IEEEproof}
To conclude the proof of Theorem \ref{theorem_real_lower_multi-block}, we use Lemma \ref{inf_lemma_multi-block} with $q=\Delta+L$, $s=2r$. \hfill \IEEEQED

 
\subsection{Proof of Theorems \ref{theorem_quaternion_upper_multi-block} and \ref{theorem_quaternion_lower_multi-block} (DMT of quaternion multi-block codes)} \label{proof_multi-block_quaternion}
Suppose $n=2p$ is even. Consider a multi-block lattice $\mathcal{L} \subset M_{n/2}(\H)^{k}$ of dimension $d=n^2 k$, and a multi-block code $\mathcal{C}(\rho)=\rho^{-\frac{r}{n}} \mathcal{L}(\rho^{\frac{r}{n}})$.
Every codeword is of the form $X=[X^{(1)},\ldots,X^{(k)}] \in \mathcal{C}(\rho)$.\\ 
Referring back to the channel model (\ref{multi-block_channel}), for all $l=1,\ldots,k$ we can write 
$$Y_c^{(l)}=\begin{pmatrix}Y_{1}^{(l)} & Y_{2}^{(l)} \end{pmatrix}, \quad H_c^{(l)}=\begin{pmatrix} H_{1}^{(l)} &  H_{2}^{(l)}\end{pmatrix}, \quad W_c^{(l)}=\begin{pmatrix}W_{1}^{(l)} & W_{2}^{(l)}\end{pmatrix},$$
where $Y_{1}^{(l)}, Y_{2}^{(l)}, H_{1}^{(l)}, H_{2}^{(l)}, W_{1}^{(l)}, W_{2}^{(l)} \in M_{m \times p}(\C)$, and we have the equivalent quaternionic channel:
\begin{small}
$$\underbrace{\begin{pmatrix} Y_{1}^{(l)} & Y_{2}^{(l)} \\ -(Y_{2}^{(l)})^* & (Y_{1}^{(l)})^* \end{pmatrix}}_{\text{\normalsize{$Y^{(l)}$}}} =\sqrt{\frac{\rho}{n}} \underbrace{\begin{pmatrix} H_{1}^{(l)} & H_{2}^{(l)} \\ -(H_{2}^{(l)})^* & (H_{1}^{(l)})^* \end{pmatrix}}_{\text{\normalsize{$H^{(l)}$}}} \underbrace{\begin{pmatrix} A^{(l)} & -(B^{(l)})^* \\ B^{(l)} & (A^{(l)})^* \end{pmatrix}}_{\text{\normalsize{$X^{(l)}$}}} + \underbrace{\begin{pmatrix} W_{1}^{(l)} & W_{2}^{(l)} \\ -(W_{2}^{(l)})^* & (W_{1}^{(l)})^* \end{pmatrix}}_{\text{\normalsize{$W^{(l)}$}}}.$$
\end{small}

\subsubsection{Proof of Theorem \ref{theorem_quaternion_upper_multi-block}}
We can write the outage probability as 
$$P_{\out}(R)=\mathbb{P}\left\{ \frac{1}{k} \left(\sum_{l=1}^{k} \log\det(I+\rho (H^{(l)})^{\dagger}H^{(l)})\right) \leq 2R\right\}.$$
Define $L=\min(m,p)$, $\Delta=\abs{p-m}$, and let $\lambda_1^{(l)} \geq \cdots \geq \lambda_{L}^{(l)},\quad l=1,\ldots,k$
the ordered nonzero eigenvalues of $(H^{(l)})^{\dagger}H^{(l)}$ with distribution 
\begin{align*}
&p(\lambda_1^{(l)},\ldots,\lambda_{L}^{(l)})= K \prod_{i=1}^{L} (\lambda_i^{(l)})^{2\Delta+1} e^{-\sum_{i=1}^{L} \lambda_i^{(l)}} \prod_{i < j} \left(\lambda_i^{(l)} - \lambda_j^{(l)}\right)^4, \quad l=1,\ldots,k.
\end{align*}
Let $\lambda_i^{(l)}=\rho^{-\alpha_i^{(l)}} \; \forall l=1,\ldots,k$, and 
$\mathcal{A}=\left\{ \boldsymbol{\alpha} \in \R^k \; :\; 
 0 \leq \alpha_1^{(l)} \leq \cdots \leq \alpha_{L}^{(l)} \;\; \forall l=1,\ldots,k
\right\}$.
Then
$$p(\boldsymbol\alpha) \doteq \rho^{-2\sum\limits_{l=1}^k\sum\limits_{i=1}^L (\Delta+1)\alpha_i^{(l)}}e^{-\sum\limits_{l=1}^k \sum\limits_{i=1}^L \rho^{-\alpha_i^{(l)}}} \prod_{l=1}^k \prod_{i<j} \left(\rho^{-\alpha_i^{(l)}}-\rho^{-\alpha_j^{(l)}}\right)^4.
$$
We have $P_{\out}(R)=\mathbb{P}\left\{ \prod\limits_{l=1}^{k} \prod\limits_{i=1}^{L} (1+\rho^{1-\alpha_i^{(l)}}) \leq \rho^{rk}\right\}\geq \mathbb{P}(\mathcal{A}_0)$,
where 
\begin{align*}
& \mathcal{A}_0=\left\{ \boldsymbol\alpha \in \mathcal{A} \;:\; \sum_{l=1}^{k} \sum_{i=1}^{L} (1-\alpha_i^{(l)})^+ \leq rk\right\}
=\left\{ \boldsymbol\alpha \in \mathcal{A} \;:\; \forall \mathbf{j}=(j_1,\ldots,j_k) \,:\, j_l \leq L \,\forall l, \; \sum_{l=1}^{k} \sum_{i=1}^{j_l} (1-\alpha_i^{(l)}) \leq rk\right\} 
\end{align*}
Given $\delta >0$, and letting $ \mathcal{S}_{\delta}=\left\{ \boldsymbol\alpha \in \mathcal{A} \;:\; \forall i \neq j, \;
\abs{\alpha_i^{(l)} - \alpha_j^{(l)}} > \delta \;\;  \forall l=1,\ldots,k \right\}$, we have the lower bound
\begin{align*}
&P_{\out}(R) \dotgeq 
\int_{\mathcal{A}_0 \cap S_{\delta}} \rho^{-2\sum\limits_{l=1}^{k} \sum\limits_{i=1}^{L} \alpha_i^{(l)} (\Delta+1) } e^{-\sum\limits_{l=1}^{k} \sum\limits_{i=1}^{L} \rho^{-\alpha_i^{(l)}}} 
\prod_{l=1}^{k} \prod_{1 \leq i<j \leq L} \left(\rho^{-\alpha_i^{(l)}}-\rho^{-\alpha_j^{(l)}}\right)^4 d\boldsymbol\alpha  \\
&\dotgeq \frac{(1-\rho^{-\delta})^{2L(L-1)k}}{e^{Lk}}\int_{\mathcal{A}_0 \cap S_{\delta}} \rho^{-2 \sum\limits_{l=1}^{k} \sum\limits_{i=1}^{L} (\Delta + 2 L - 2i +1) \alpha_i^{(l)}}d\boldsymbol\alpha.
\end{align*}     
Using Lemma \ref{inf_lemma_multi-block} with  $q=\Delta +L$, $s=r$, we find that the DMT upper bound is $2\inf_{\boldsymbol\alpha \in \mathcal{A}_0} F(\boldsymbol\alpha)=k d_2(r)$. \hfill \IEEEQED

\subsubsection{Proof of Theorem \ref{theorem_quaternion_lower_multi-block}}
We only highlight the main steps of the proof. \\
Letting $H=\diag(H^{(1)},\ldots,H^{(k)})$ and  $W=[W^{(1)},\ldots,W^{(k)}]$ the multi-block quaternion channel matrix and noise, we have $P_e(H) \leq \mathbb{P}\{ \norm{W}^2 > d_H^2/4\},$
with
$$d_H^2=\frac{\rho}{n} \min_{\substack{X, X' \in \mathcal{C}(\rho) \\ X \neq X'}} \sum_{l=1}^k \norm{H^{(l)}(X^{(l)}-X'^{(l)})}^2 \geq \frac{1}{p} \rho^{1-\frac{2r}{n}} \sum_{l=1}^k \sum_{i=1}^L \lambda_i^{(l)} \mu_i^{(l)},$$
where 
$0 \leq \mu_1^{(l)}={\mu_1^{(l)}}' \leq \cdots \leq \mu_p^{(l)}={\mu_p^{(l)}}'$ are the ordered eigenvalues of $\Delta X^{(l)} (\Delta X ^{(l)})^{\dagger}$ with $\Delta X=X-X'$. \\
For any $\mathbf{j}=(j_1,\ldots,j_k)$ with $J=\sum_{l=1}^k j_l \geq 1$, we have
\begin{align*}
d_H^2 \geq 
\frac{J}{p} \rho^{1-\frac{2r}{n}} \left( \prod_{l=1}^k \prod_{i=1}^{j_l} \lambda_i^{(l)} \mu_i^{(l)}\right)^{\frac{1}{J}}.
\end{align*}
Note that $\forall i=1,\ldots, p$, $\forall l=1,\ldots,k$, $\mu_i^{(l)} \dotleq \rho^{\frac{2r}{n}}$, and 
$$\prod_{l=1}^k \prod_{i=1}^{j_l} \mu_i^{(l)} =\frac{\det(\Delta X \Delta X^{\dagger})^{\frac{1}{2}}}{\prod_{l=1}^k \prod_{i=j_l+1}^p \mu_i^{(l)}} \geq \frac{1}{4^{kp-J}\rho^{(kp-J)\frac{r}{p}}}.$$
Therefore $\forall \mathbf{j} \neq 0$, $d_H^2 \geq c_j rho^{\delta_{\mathbf{j}}(\boldsymbol\alpha,r)}$, 
where $\delta_{\mathbf{j}}(\boldsymbol\alpha,r)=-\frac{1}{\sum_{l=1}^k j_l}\sum\limits_{l=1}^k \left(\sum\limits_{i=1}^{j_l} \alpha_i^{(l)} + r -j_l\right)$, and $c_j$ is a suitable constant. We have
$$P_e \leq \int \mathbb{P} \left\{ \norm{W}^2 > \frac{d_H^2}{2}\right\} p(\boldsymbol \alpha) d \boldsymbol\alpha.$$
The distribution $p(\boldsymbol\alpha)$ in (\ref{product_distribution}) is upper bounded by 
$$p(\boldsymbol\alpha) \dotleq p'(\boldsymbol\alpha)=e^{-\sum\limits_{l=1}^k \sum\limits_{i=1}^L \rho^{-\alpha_i^{(l)}}} \rho^{-\sum\limits_{l=1}^k \sum\limits_{i=1}^L \alpha_i^{(l)} N_i}$$
where $N_i=2(\Delta+ 2L - 2i +1)$. \\
Since $2\norm{W}^2 \sim 2\chi^2(2mpk)$, we have 
$$P_e \leq  \int_{\mathcal{A}} p'(\boldsymbol\alpha) \Phi_{mpk}\left(\frac{d_H^2}{2}\right) d\boldsymbol\alpha.$$
To conclude the proof, we use Lemma \ref{laplace_lemma_multi-block} and Lemma \ref{inf_lemma_multi-block} with $q=\Delta+L$, $s=r$. \hfill \IEEEQED

\begin{footnotesize}
\bibliographystyle{IEEEtran}

\bibliography{STC}

\end{footnotesize}

\end{document}